\documentclass[prb,twocolumn,showpacs,superscriptaddress]{revtex4}
\usepackage{graphicx}
\usepackage[T1]{fontenc}
\usepackage{amssymb,amsmath}
\usepackage{dcolumn}

\begin{document}

\title{Phonons and  the coherence scale of models of heavy fermions}

\author{     M. Raczkowski}  
\affiliation{Institut f\"ur Theoretische Physik und Astrophysik, 
             Universit\"at W\"urzburg, Am Hubland, D-97074 W\"urzburg, Germany}
\affiliation{Marian Smoluchowski Institute of Physics, 
             Jagellonian University, Reymonta 4, PL-30059 Krak\'ow, Poland}

\author{     P. Zhang} 
\affiliation{Institut f\"ur Theoretische Physik und Astrophysik, 
             Universit\"at W\"urzburg, Am Hubland, D-97074 W\"urzburg, Germany}
\affiliation{Department of Physics and Astronomy, Louisiana State University,
             Baton Rouge LA 70803, USA }

\author{     F. F. Assaad} 
\affiliation{Institut f\"ur Theoretische Physik und Astrophysik, 
             Universit\"at W\"urzburg, Am Hubland, D-97074 W\"urzburg, Germany}

\author{     T. Pruschke} 
\affiliation{Institute for Theoretical Physics, University of G\"ottingen, 
             Friedrich-Hund-Platz 1, D-37077 G\"ottingen, Germany}

\author{     M. Jarrell}
\affiliation{Department of Physics and Astronomy, Louisiana State University,
             Baton Rouge LA 70803, USA }

\date{\today}

\begin{abstract}
We consider  models of heavy fermions in the strong coupling or local moment limit  and include phonon degrees of 
freedom on the conduction electrons. Due to the large mass or low coherence
temperature of the heavy fermion state, it is shown that such a regime is dominated by vertex
corrections which leads to the  complete failure of the Migdal theorem. 
Even at weak electron-phonon couplings, binding of the conduction
electrons  competes with the Kondo effect  and substantially 
reduces  the coherence temperature,  ultimately leading to the Kondo breakdown.  Those results are obtained 
using  a combination of the slave boson method and  Migdal-Eliashberg  approximation as well as the
dynamical mean-field theory approximation.  
\end{abstract}

\maketitle

\section { \label{sec:0} Introduction and  Models }  

The heavy fermion paramagnetic state as realized for instance in CeCu$_6$ or
induced by a weak magnetic field in YbRh$_{2}$(Si$_{1-x}$Ge$_{x}$)$_2$ 
corresponds to  the  coherent, Bloch like, superposition of the individual
Kondo screening clouds of the spins of rare-earths.\cite{lohneysen08}  The coherence temperature
$T_{\rm coh}$,  or inverse effective mass of this state is set by the Kondo
\cite{Ueda86,Georges00,Pruschke00,Assaad04a} scale which lies orders of 
magnitude below the Fermi  temperature, $T_F$,  of the host metallic state.  In general, for weakly correlated metals, 
the characteristic  phonon frequencies set by the Debye temperature  $\Theta_D$ are much smaller than the Fermi 
temperature $E_F\simeq 10^{4-5}$K thus  yielding a small parameter $\Theta_D/
T_F$  on which  Migdal theorem  is based. \cite{Migdal58}
%Strictly speaking 
This small parameter is  absent in heavy fermion materials since the  Fermi  
temperature  should be  replaced by $T_{\rm coh}$ which might be even
smaller than 1K. Moreover, smallness of the  relevant energy scales implies
that  properties of the ground state can be easily tuned by Hamiltonian perturbations. 
This observation raises the central question of this paper:  what  role  do  phonons  play in models of heavy fermions?

We address this question on the basis of the periodic Anderson model (PAM) on a square 
lattice with Holstein phonons that couple to the  conduction band electrons: 
\begin{equation}
\label{pam}
%\label{Hamiltonian}
\begin{aligned}
    H =   &  H_0  + H_V + H_{ph}  \text{ with }     \\
	H_0  = &  \sum_{ \pmb{k},\sigma} \epsilon({\pmb k}) c^{\dagger}_{ {\pmb k},\sigma}
                       c^{}_{{\pmb k}, \sigma},  \\
	H_V  = & V \sum_{\pmb i,\sigma} \bigl( f^\dagger_{ {\pmb i} ,\sigma} c^{}_{ {\pmb i},\sigma}
                      + h.c. \bigr)   + 
                 (\epsilon_f-\mu) \sum_{{\pmb i},\sigma} f^\dagger_{ {\pmb i}, \sigma} f^{}_{ {\pmb i}, \sigma}  \\
             + & U \sum_{\pmb i}  \bigl(n^f_{{\pmb i}, \uparrow}-1/2\bigr)
       \bigl(n^f_{ {\pmb i}, \downarrow}-1/2\bigr),  \\
      H_{ph} = & g \sum_{ {\pmb i} } \hat{Q}_{\pmb i}( n^c_{ {\pmb i}} -1 )  
                + \sum_{\pmb i} \Bigl( \frac{\hat{P}_{\pmb i}^2}{2M} + \frac{k}{2}
                \hat{Q}_{\pmb i}^2\Bigr).
\end{aligned}
\end{equation}
Here, $H_0$  describes the conduction band with dispersion relation $ \epsilon({\pmb k})   = -2t (\cos k_x + \cos k_y) -\mu$ and with
$c^\dagger_{{\pmb k},\sigma}$  creating a conduction electron with $z$-component of spin $\sigma$ and in the Bloch state with crystal momentum
${\pmb k}$. Next, $H_V$ accounts for the hybridization with a conduction electron in Wannier state centered around  unit cell  $ {\pmb i} $ and the
localized $f$-electron  in the same unit cell while the Coulomb repulsion set by the Hubbard $U$  on the $f$-orbitals 
accounts for local moment formation.  Finally, $H_{ph}$ corresponds to Einstein phonons  with a Holstein coupling to the 
conduction  electrons.  In fact, the choice between a model in which phonons couple predominately
either to the conduction or $f$-electrons is material dependent. Indeed,
retaining  the coupling  of Holstein phonons only to the conduction electrons
is justified in the local moment regime where charge fluctuations on the
$f$-orbitals are suppressed due to the  strong Coulomb repulsion. In this limit
the model maps onto the Kondo lattice model (KLM):
\begin{equation}
	H_{KLM} = H_0 + H_{ph} + J \sum_{\pmb i} {\pmb S}_{\pmb i}^c  {\pmb
          S}_{\pmb i}^f.  
\label{klm}
\end{equation}
In contrast, the effect of phonon coupling to the $f$-electrons is expected to become more important in the mixed
valence regime of moderately heavy fermions such as filled skutterudites as emphasized in Ref. \onlinecite{Ono05}.

\begin{figure}
\begin{center}
\includegraphics[width=0.4\textwidth]{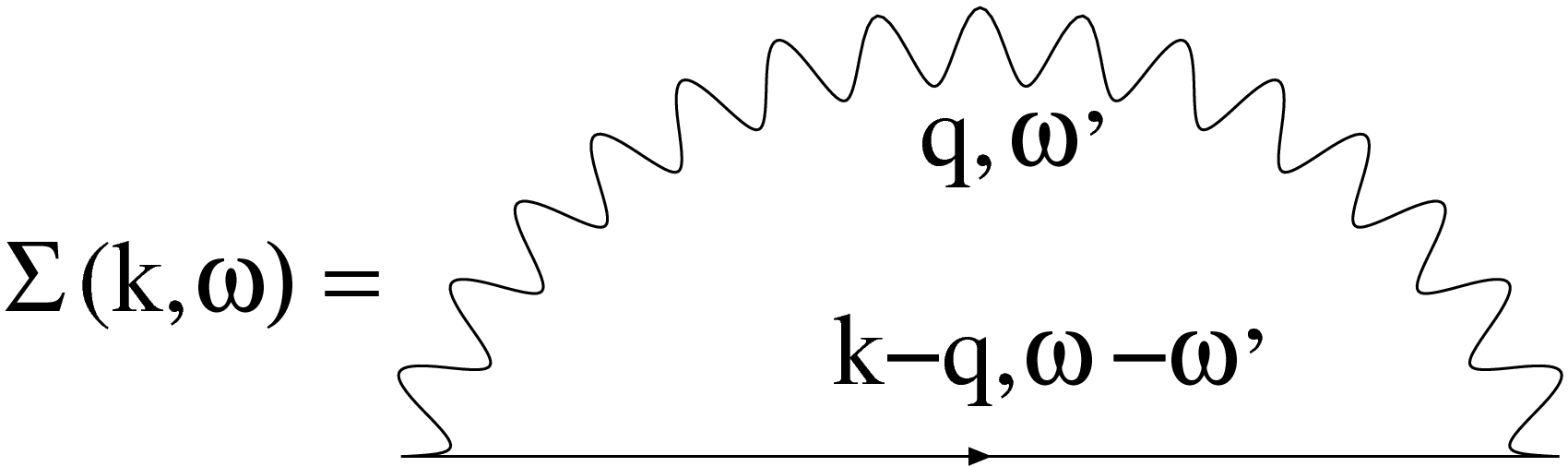}
\end{center}
\caption{ Self-energy diagram. The solid (wavy) line corresponds to the bare single-particle
Green's function (phonon propagator), respectively. }
\label{SCB.fig}
\end{figure}

We use several  methods to unravel the physics contained in those  model Hamiltonians.  
Arbitrarily low temperatures as well as phonon frequencies may be reached within a combination of the slave-boson (SB) 
mean-field approximation \cite{Kotliar86} with the  self-consistent calculation of the self-energy diagram of
Fig.~\ref{SCB.fig} within Migdal-Eliashberg \cite{Migdal58,Eliashberg60,Engelsberg63} (ME)  approximation.  
The former leads to the hybridized band picture of the heavy fermion state
while the latter allows one to account for coupling to the phonons. Furthermore, since the SB
approximation fails at finite temperatures while the ME approach does not
  capture vertex corrections, we equally use the dynamical mean-field theory (DMFT) approximation with a recently
developed weak coupling continuous-time quantum Monte Carlo (CT-QMC) impurity solver\cite{Rubtsov05} to
include  phonon degrees of freedom into the PAM. \cite{Assaad07,Assaad08}   
Finally, in order to access the low temperature limit, we resort to numerical
renormalization group  (NRG) method \cite{Bulla08} as a
complementary to the CT-QMC impurity solver applied to the KLM (\ref{klm}). 

To understand the failure of the ME approximation for the heavy fermion state,  let us start with 
the phonons coupled to a band of conduction electrons with a flat density of
states  of width $W$. In this case, the self-energy diagram of Fig. \ref{SCB.fig}  gives a mass 
renormalization,
\begin{equation}
\label{M_eff_Holstein.Eq}
	\frac{m^{*}}{m_0}  =  1 + \lambda \left( 1 - \frac{\omega_0} {W/2 + \omega_0} \right)
 \text{ where }    
\lambda =  \frac{ g^2}{k W}
\end{equation}
and $\omega_0  =  \sqrt{k/M} $ corresponds to the phonon frequency, and  $m_0$ is the  mass of the bare electron.  
$\lambda $ corresponds to the dimensionless 
electron-phonon interaction and we have explicitly  included high frequency corrections. 

In the adiabatic limit $\omega_0 << T_{\rm coh} $, we can use the above formula with the heavy Fermi liquid as a starting point 
rather than the bare conduction electrons to account for the effect of phonons.  
As it is reviewed in some details in Sec. \ref{sec:1}, the ratio of the coherence temperature of the heavy fermion state 
to the  Fermi temperature  of the host metal is   given by the quasiparticle (QP) residue, 
\begin{equation}
	Z = | \langle \Psi ^{N-1}_0 | c_{{\pmb k}_F,\sigma}   | \Psi ^{N}_0 \rangle  |^2. 
\end{equation} 
This quantity  measures the overlap  of the QP with a bare conduction electron. Hence, the coupling of the 
phonons to the  heavy fermion quasiparticles will be renormalized by a factor $Z$ in comparison to the coupling 
to the bare electron (i.e. $g \rightarrow g Z$).  One equally expects effective bandwidth to be scaled by the same factor Z.   
Accounting for these  two renormalization factors, and neglecting the high
frequency correction in Eq. (\ref{M_eff_Holstein.Eq}) gives a mass enhancement,
\begin{equation}
\label{mstar_intro.eq}
	\frac{ m^{*} }{m}  \simeq  1 + \lambda Z,  
\end{equation}
where $m \simeq m_0/Z $ corresponds to the effective mass of the heavy fermion
state in the absence of phonons. Therefore, in this limit the coherence scale,
\begin{equation}	
	T_{\rm coh} \propto \frac{m_0}{m^{*}} \simeq   \frac{Z}{1  + \lambda Z}, 
\end{equation}
is next to unaffected by the inclusion of phonons since $Z << 1$ even in  the strong electron-phonon coupling limit, 
$\lambda \simeq 1$. In Sec. \ref{sec:1} we verify explicitly this result  with the use  of a combination 
of  the SB technique and the ME approximation. In this case the coherence temperature is also protected from  
{\it adiabatic} phonons which stems from their coupling to merely a fraction, $ Z$, of the heavy QP. 
  
In the high frequency limit,  $ W  >> \omega_0 >>  T_{coh} $ the correct starting point is to consider the 
Hamiltonian $H_0 + H_{ph} $.  Neglecting vertex corrections, one can account for the low energy physics of this
 Hamiltonian  within the ME approximation.  This approximation describes the      
formation of  quasiparticles with enhanced effective mass set by $m_0 ( 1 + \lambda )$ or equivalently a bandwidth reduced by 
a factor $(1 +  \lambda)$.  From there onwards one can account for the magnetic impurities. The hybridization  
matrix element between the  QP  and  the $f$-electron is  
renormalized by a factor $1/\sqrt{1 +  \lambda }$ in comparison to the hybridization with the bare electron.  
Taking into account those two renormalization factors, bandwidth and hybridization,  the Kondo temperature, 
$ T_K \propto \exp ( -  U W/4V^2 )$, remains invariant at this level of
approximation. Hence, as in the adiabatic  case, one also expects in this regime a very weak influence of  the  phonon degrees
of freedom on the scales of the heavy fermion state.  
This point is confirmed explicitly in Sec. \ref{sec:1} within the ME approximation.

Therefore, we anticipate that retaining only non crossing diagrams thus omitting vertex corrections  leaves the scales 
of the heavy fermion state  next to unaffected.  Vertex corrections will
clearly play a role since they  lead to  polaron binding which competes with the Kondo screening of the impurity spins.   
In particular, integrating out the phonon degrees of freedom yields a retarded attractive interaction.  
The retardation is set  by $1/\omega_0$ and its magnitude by $\lambda W/2 $.  This term  leads  to  binding between  polarons 
into singlets and is responsible for the onset of superconductivity.  
To capture this competition and associated  substantial  reduction of the coherence temperature even at very low values of the 
electron phonon coupling, we carry out in Sec. \ref{sec:2} DMFT calculations 
using two complementary impurity solvers based on: (i) a recently developed CT-QMC technique\cite{Rubtsov05,Assaad07} as well as
(ii) zero-temperature NRG method.\cite{Bulla08}  
Finally, in Sec. \ref{sec:3} we conclude and discuss the implications of our results.
 
Unless stated otherwise, throughout the paper  we  consider  the parameters
$U/t =4$, $V/t=1$, and $\epsilon_f = \mu$  for the PAM and $J=W/2$ with $W=0.2$ for the KLM, and
choose  the chemical potential $\mu$ such that the total particle number per unit cell reads $\langle n  \rangle =1.8$.

\section{\label{sec:1} Mean-field approximation }  

In order to understand the basic underlying physics of the model
Eq.~(\ref{pam}) it is instructive to work first in the mean-field framework. 
A good starting point to handle strong electron correlations is offered by the
SB mean-field method. \cite{Kotliar86}  Indeed, its usefulness has been proved in 
the studies devoted to searching for the ground-state of both the doped Hubbard 
model \cite{Raczkowski06} and  its multi-band variant. \cite{Fresard97} Moreover, it has been 
applied to determine instabilities of the PAM, \cite{pam_mag} 
and since it captures two characteristic energy scales, i.e., the coherence $T_{\rm coh}$ and 
the Kondo $T_{\rm K}$ ones, is believed to account for the essential physics of the heavy-fermion 
systems. \cite{Burdin09} 

Additionally, in order to gain  preliminary insight into the interplay between electron
correlations and electron-phonon coupling,  we  combine the SB technique
with the ME approximation. \cite{Migdal58,Eliashberg60,Engelsberg63} 
%The latter involves summing over noncrossing diagrams contributing the electron self-energy  by self-consistently solving the diagram 
%of Fig. \ref{SCB.fig}. 
Remarkably, despite obvious flaws there are some  regimes  where this 
approach works quite well \cite{Brunner00b,Mishchenko01b,Luscher06} and on top
it might be systematically improved by including the
leading-order vertex corrections in the self-energy calculations. 
For example, it has widely been used to describe the dynamics of a single hole in the 
magnetically ordered background of Mott insulators and to analyze changes in the QP properties 
by coupling to various bosonic excitations such  as  magnons, \cite{Martinez91,Ramsak98} 
phonons, \cite{Slezak06,Gunnarsson06} and orbitons. \cite{Brink00} 
Moreover, recent studies of the one-dimensional quarter-filled Holstein model 
have shown that in the temperature range above the onset of the Luttinger
liquid phase, the approximation  is capable of reproducing the temperature dependence of 
the one-particle spectral function obtained within a cluster extension of the DMFT. \cite{Assaad08}

\subsection{\label{sec:1c}  Energy scales in the PAM}

\begin{figure}[t!]
\begin{center}
\includegraphics[width=0.45\textwidth]{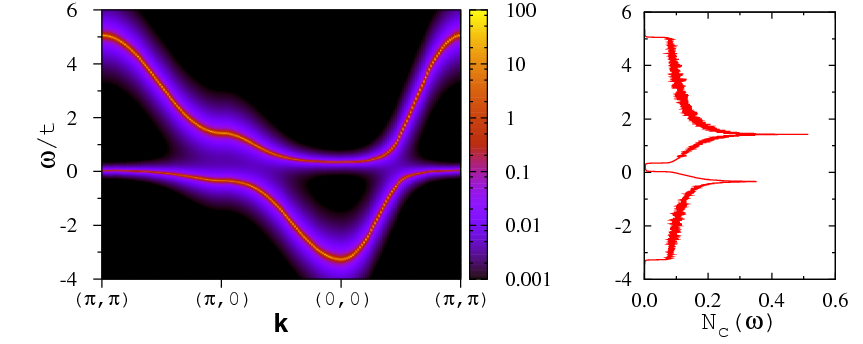}
\end{center}
\caption
{(Color online) $c$-electron spectral function $A_c({\bf k},\omega)$ (left)
  and the corresponding density of states $N_c(\omega)$ obtained in the PAM with
  $V/t=1$ in the coherent Fermi liquid regime at the low temperature $\beta t= 400$.
}
\label{AN_full}
\end{figure}

%We begin by discussing  energy scales in the PAM that separate ground states
%with different properties. 
It is now well established that both the PAM and its
strong coupling version, i.e., the KLM have two paramagnetic 
solutions. The first one sets in above the scale $T_{\rm K}$ and is
characterized by the fully localized $f$-electrons. In contrast,  below the Kondo
temperature $T_{\rm K}$, the $c$- and $f$-electrons couple to form a
heavy fermion state that on decreasing $T$ evolves eventually 
into a coherent Fermi liquid state and the onset is marked by the so-called
coherence scale  $T_{\rm coh}$. Its most prominent features are: 
(i) a flat low-intensity region in the $c$-electron spectral function $A_c({\bf
  k},\omega)$ that results from the hybridization between a dispersionless
correlated $f$-band and a wide conduction $c$-band; as shown in
Fig.~\ref{AN_full} it crosses the Fermi level and hence it is responsible for a metallic 
character of the low-temperature ground-state; 
(ii) quasiparticles with a strongly renormalized effective mass $m/m_0\gg 1$, 
and (iii) linear specific heat $C_v=\gamma T$. 

\begin{figure}[b!]
\begin{center}
\includegraphics[width=0.45\textwidth]{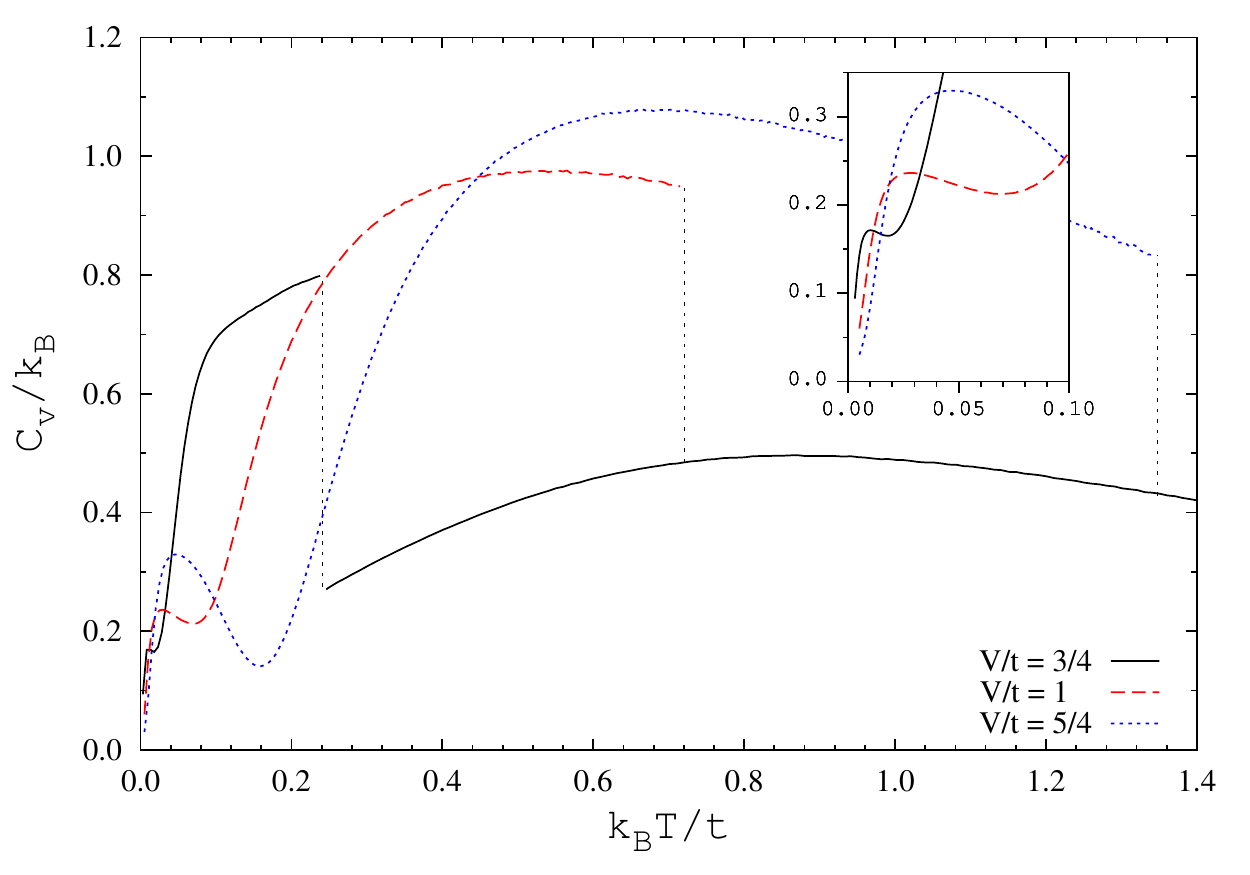}
\end{center}
\caption
{(Color online) Temperature dependence of the specific heat $C_v$: 
discontinuity of $C_v$  in the high-temperature regime signals  Kondo
breakdown at $T_{\rm K}$; low-temperature peaks  indicate onset of the Fermi liquid regime at
$T_{\rm coh}$. Inset shows a closeup of the low-temperature limit.   
}
\label{c_v}
\end{figure}

Within the SB approach one can define a Kondo temperature $T_{\rm K}$. It corresponds to the temperature scale at which the SB
factor $z$  renormalizing the hybridization amplitude $V$  vanishes. 
The transition into the state with singly occupied $f$-orbitals at $T_{\rm K}$ is also seen
as discontinuity of the specific heat $C_v=-T\bigl(\tfrac{\partial^2 F_{SB}}{\partial
  T^2}\bigr)_{\langle n\rangle}$ shown in Fig.~\ref{c_v}. The second  feature of $C_v$ is a
low-temperature peak at $T_{\rm coh}$ that signals entering the coherent
Fermi liquid regime with well defined QP peaks whose effective mass is
strongly enhanced. The two energy scales  $T_{\rm coh}$ and
$T_{\rm K}$ read off from Fig.~\ref{c_v} are listed in Table~\ref{tab_SB}.
As expected based on the Gutzwiller approach, \cite{Ueda86} large $N$-method,
\cite{Georges00}  DMFT studies, \cite{Pruschke00} as well as on the QMC
simulations,\cite{Assaad04a} one finds
that both energy scales  track  each other on varying the hybridization
amplitude $V$. To be more precise  within the large $N$ approach,  $T_{\rm coh} $ lies well below the Kondo temperature 
with proportionality constant being  strongly dependent on the conduction density of states.   However, at 
fixed density and upon varying the hybridization,   
$T_{\rm coh}$  tracks $ T_{\rm K}$. In particular, for a flat
density of states, the  Gutzwiller method yields 
$T_{\rm K}\propto \exp(-UW/4V^2)$ and accordingly increasing $V$ shifts
both characteristic temperatures towards higher $T$. Remarkably, even though
$T_{\rm K}$ changes by one order of magnitude on increasing $V/t$ from $3/4$
to $5/4$,  the ratio $T_{\rm coh}/T_{\rm K}$ is next to constant  ({\it cf.} Table~\ref{tab_SB}).

\begin{figure}[t!]
\begin{center}
\includegraphics[width=0.45\textwidth]{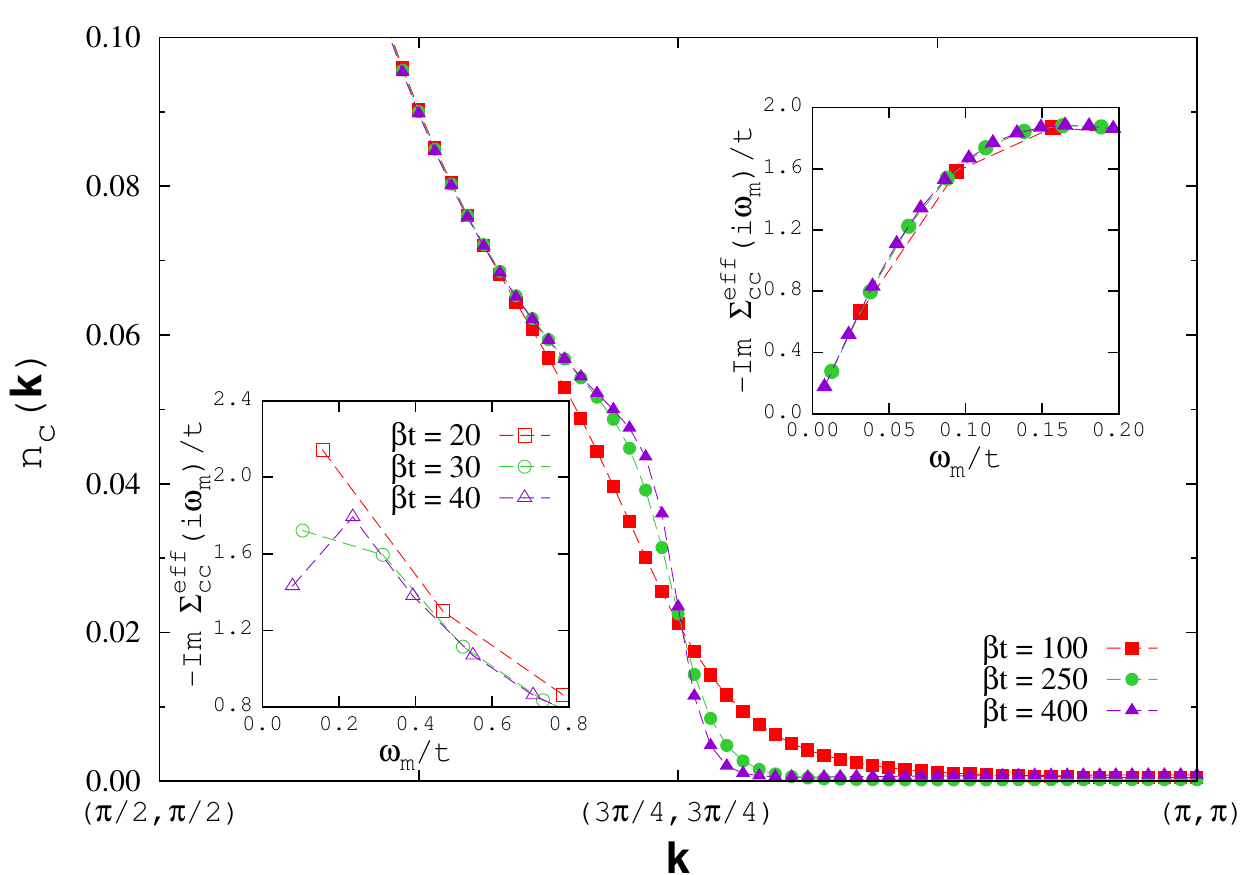}
\end{center}
\caption
{(Color online) $c$-electron single-particle occupation $n_{c}({\bf k})$ along
the nodal direction of the Brillouin zone obtained in the PAM with
$V/t=1$. Insets show the imaginary part of the effective $c$-electron self-energy $-{\rm Im}
\Sigma_{cc}^{eff}(i\omega_m)$  as a function of Matsubara frequencies
$\omega_m$. 
}
\label{S_ef}
\end{figure}

%%%%%%%%%%%%%%%%%%%%%%%%%%%%%%%%%%%%%%%%%%%%%%%%%%%%%%%%%%%%%%%%%%%%%%%%
%%                              table 1
%%%%%%%%%%%%%%%%%%%%%%%%%%%%%%%%%%%%%%%%%%%%%%%%%%%%%%%%%%%%%%%%%%%%%%%%
\begin{table}[b!]
\caption {
Slave-boson coherence temperature $T_{\rm coh}$, Kondo temperature $T_{\rm
  K}$, their ratio, as well as QP residue $Z$ obtained from Eq. (\ref{Z}) in the PAM at $\beta
t=400$ for a few representative values of the hybridization $V$.
}    
\begin{ruledtabular}
\begin{tabular}{cddd}
 $V/t$                &     0.75   &    1.0    &    1.25     \cr
\colrule
$k_BT_{\rm coh}/t$      &     0.010  &    0.026  &    0.046    \cr
$k_BT_{\rm K}/t$        &     0.244  &    0.721  &    1.346    \cr
$T_{\rm coh}/T_{\rm K}$  &     0.041  &    0.036   &   0.034     \cr  
 $Z$                  &     0.018  &    0.043   &   0.066   \cr 
\end{tabular}
\end{ruledtabular}
\label{tab_SB}
\end{table}

The onset of the coherent Fermi liquid is also signalized by the imaginary
part of the effective $c$-electron self-energy $-{\rm Im}
\Sigma_{cc}^{eff}(i\omega_m)$ (see Appendix). Indeed, as shown in the inset in Fig.~\ref{S_ef} with
$V/t=1$, a conspicuous change in the behavior of $-{\rm Im}
\Sigma_{cc}^{eff}(i\omega_m)$ with $\omega_m\to 0$  takes place in the
temperature region $30<\beta t<40$ matching the low-temperature peak in
$C_v$. 
Moreover, a discontinuity in the $c$-electron single-particle
occupation $n_{c}({\bf k})=\int_{-\infty}^{\mu}d\omega A_c({\bf k},\omega)$
emerges below $\beta t=100$ (see Fig.~\ref{S_ef}). In fact,
since the coherent QP part of the spectral function is given at low temperature
by a delta function of weight $Z$: $A_c({\bf k},\omega)\sim
Z\delta(\omega-\varepsilon_{\bf k})$, its magnitude is given precisely by the QP residue
$Z$ which in turn is directly related to the effective QP mass $Z^{-1}=m/m_0$. 
In the low temperature limit, the value of $Z$ can be  determined as the slope of the imaginary part of the
effective self-energy on the Matsubara axis (see the inset in Fig.~\ref{S_ef}): 
\begin{equation}
Z = 
\left[1 - \frac{ {\rm Im} \Sigma_{cc}^{eff}(i\omega_m) }
{ \omega_m} \right]^{-1}_{\omega_m = \pi T  }. 
\label{Z}
\end{equation}

\begin{figure}[t!]
\begin{center}
\includegraphics[width=0.45\textwidth]{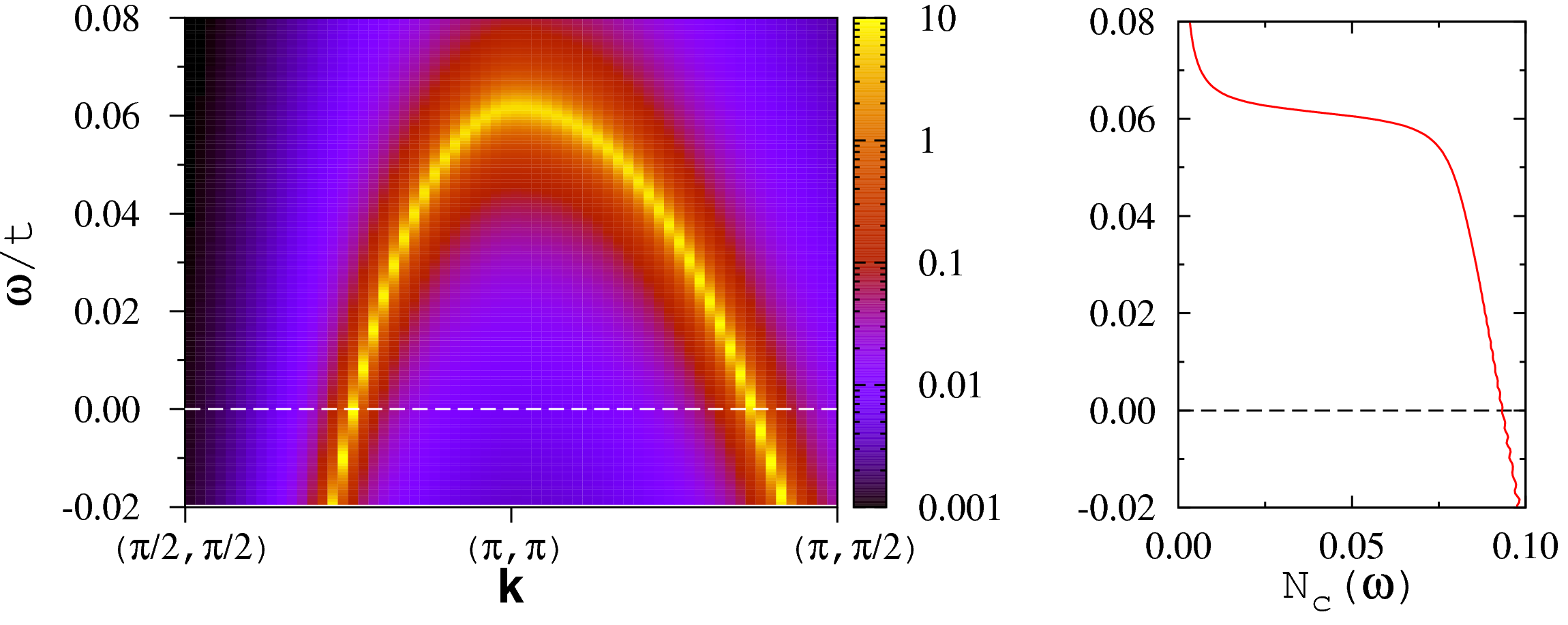} \\
\includegraphics[width=0.45\textwidth]{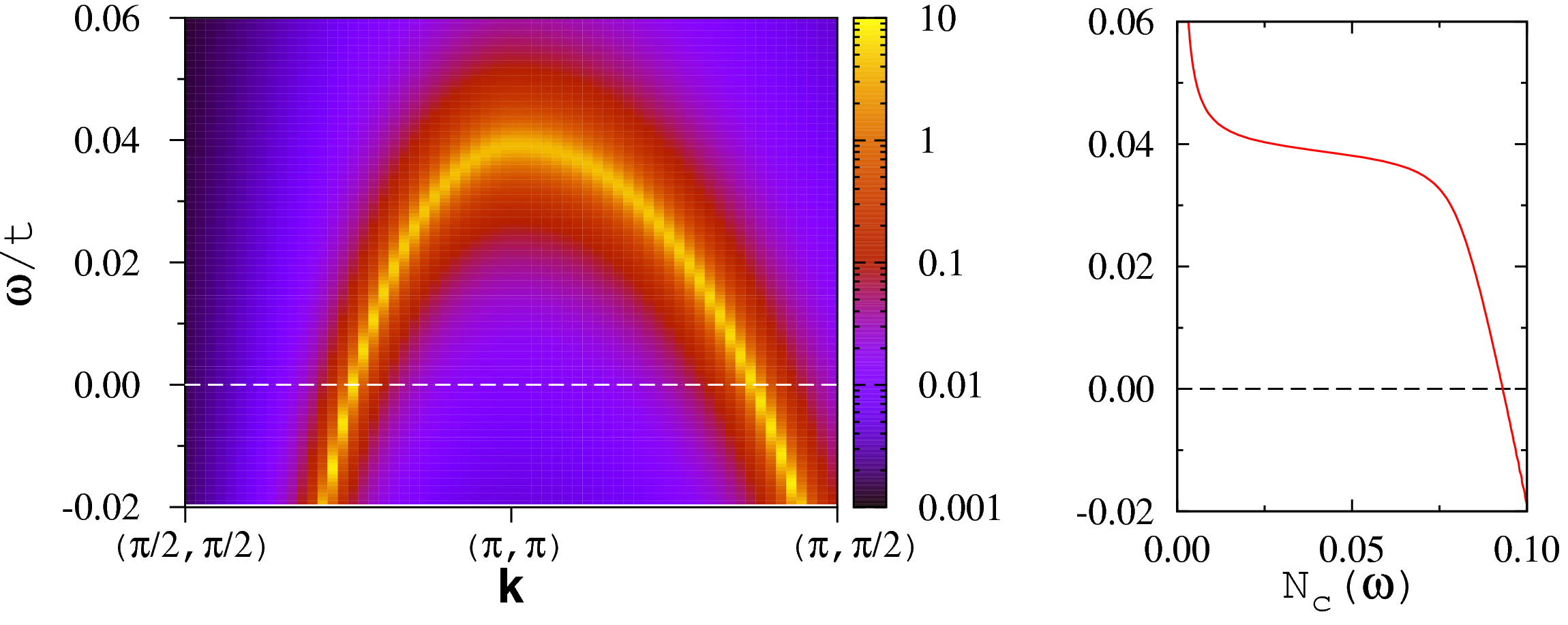}   \\
\includegraphics[width=0.45\textwidth]{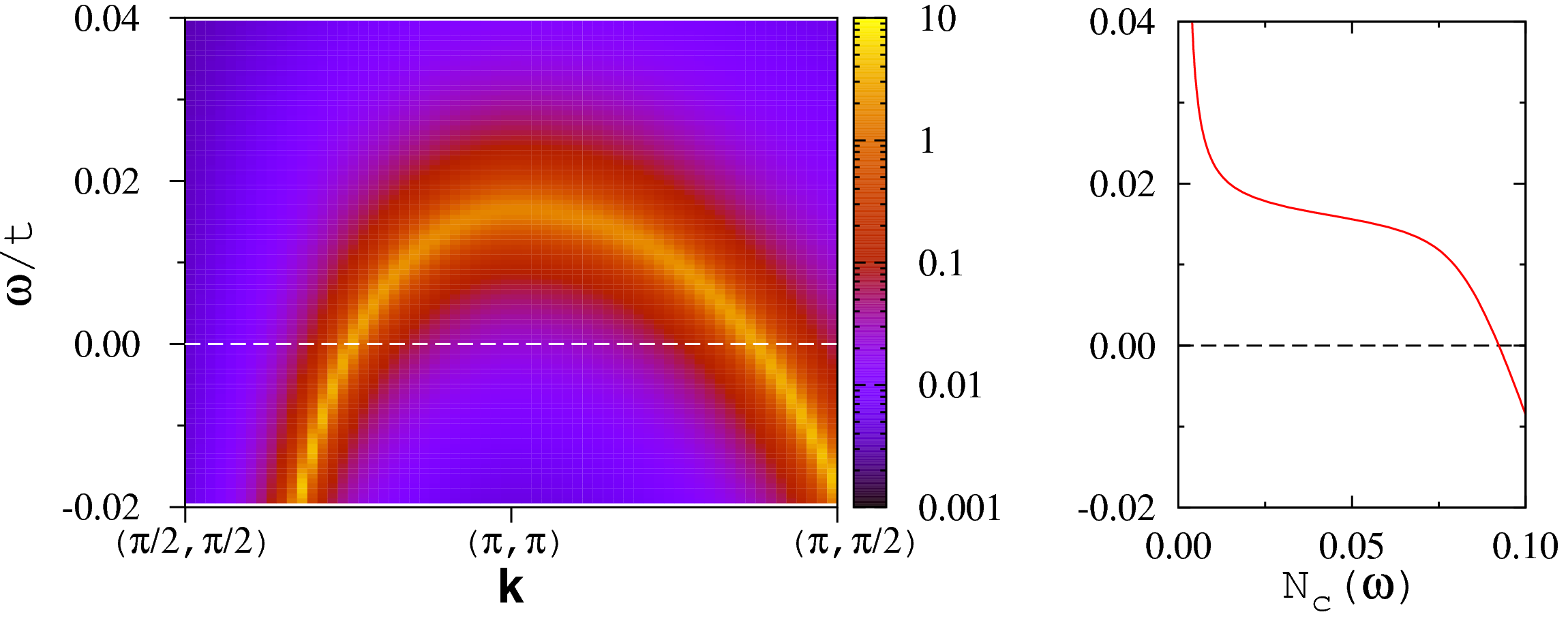}
\end{center}
\caption
{(Color online) Low-energy part of the $c$-electron spectral function $A_c({\bf k},\omega)$ (left)
  and the corresponding density of states $N_c(\omega)$ (right) in the PAM 
  found at $\beta t= 400$: $V/t=1.25$ (top), $V/t=1$ (middle), and $V/t=0.75$ (bottom).   
}
\label{AN}
\end{figure}

Furthermore, we observe that the value of $V$ strongly influences the shape and the intensity of the QP band 
around the Fermi level depicted in Fig.~\ref{AN}. On the one hand, $V/t=5/4$ yields a
more dispersive QP band with a higher intensity, as compared to the $V/t=1$
case. It extends much above the Fermi energy up to a certain value $\omega_c/t\simeq 0.06$. 
On the other hand, a smaller  $V/t=3/4$ produces a strongly renormalized
low-intensity flat band just above the Fermi level extending up to $\omega_c/t\simeq
0.02$.   Hence, in the
considered range of the hybridization amplitude $V$, the SB results indicate
that both energy scales  $T_{\rm coh}$ and $\omega_c$  are intimately
related, i.e., $T_{\rm coh}\sim\omega_c$.

\subsection{\label{sec:1d} Effect of electron-phonon coupling} 

\begin{figure}[t!]
\begin{center}
\includegraphics[width=0.45\textwidth]{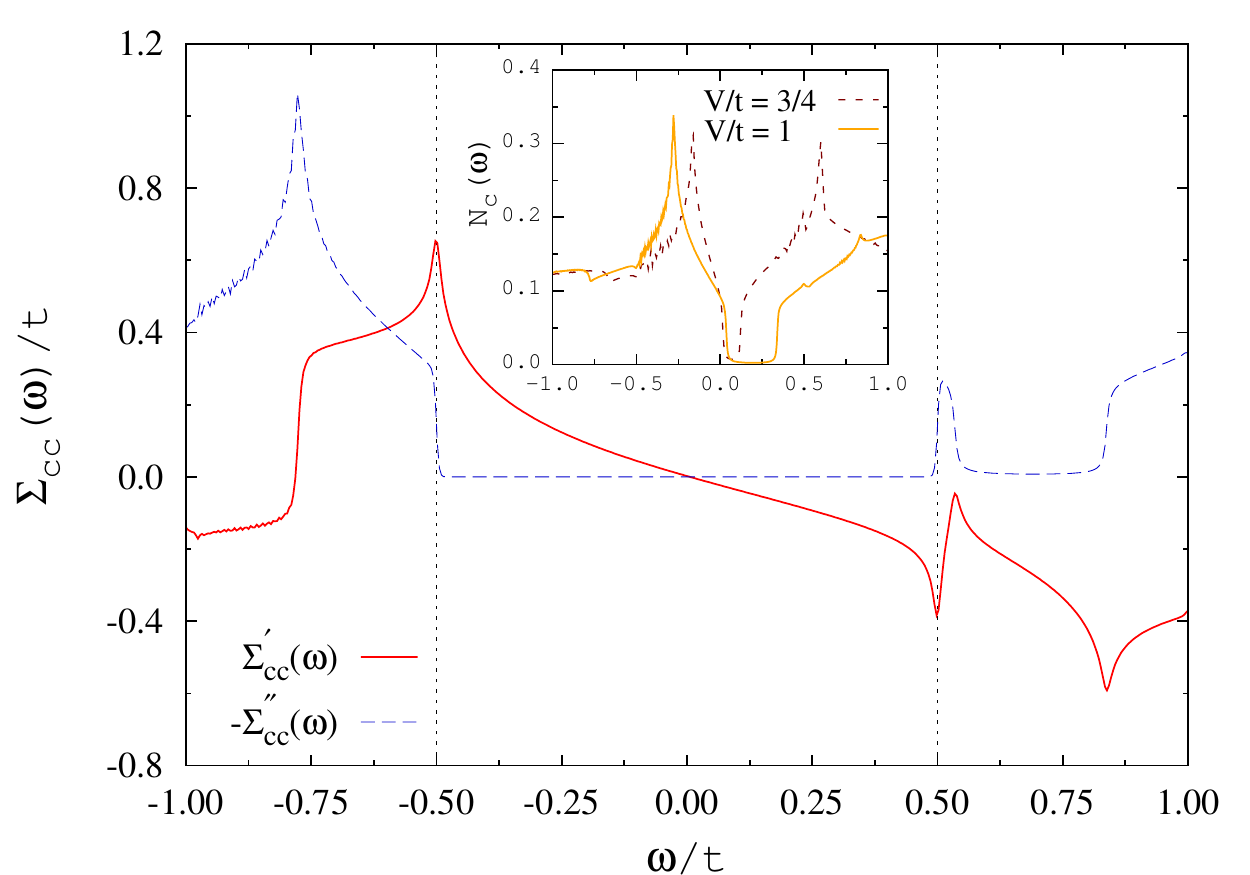}
\end{center}
\caption
{(Color online) Low-energy part of the real $\Sigma_{cc}'(\omega)$ and imaginary
  $-\Sigma_{cc}''(\omega)$ part of the $c$-electron self-energy obtained 
  with $V/t=1$, $\lambda=0.5$, and $\omega_0/t=0.5$ at $\beta t=400$. 
  Inset depicts the corresponding $c$-electron density of states $N_c(\omega)$ and, for comparison,  
 $N_c(\omega)$ obtained with $V/t=3/4$.  
}
\label{re_im}
\end{figure}

\begin{figure}[b!]
\begin{center}
\includegraphics[width=0.45\textwidth]{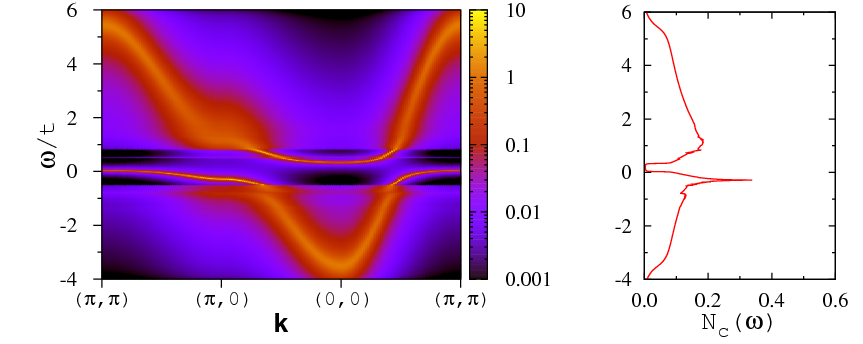}
\end{center}
\caption
{ (Color online) Same as in Fig.~\ref{AN_full} but with $\lambda=0.5$ and $\omega_0/t=0.5$.
} 
\label{AN_full_ph}
\end{figure} 

\begin{figure}[t!]
\begin{center}
\includegraphics[width=0.45\textwidth]{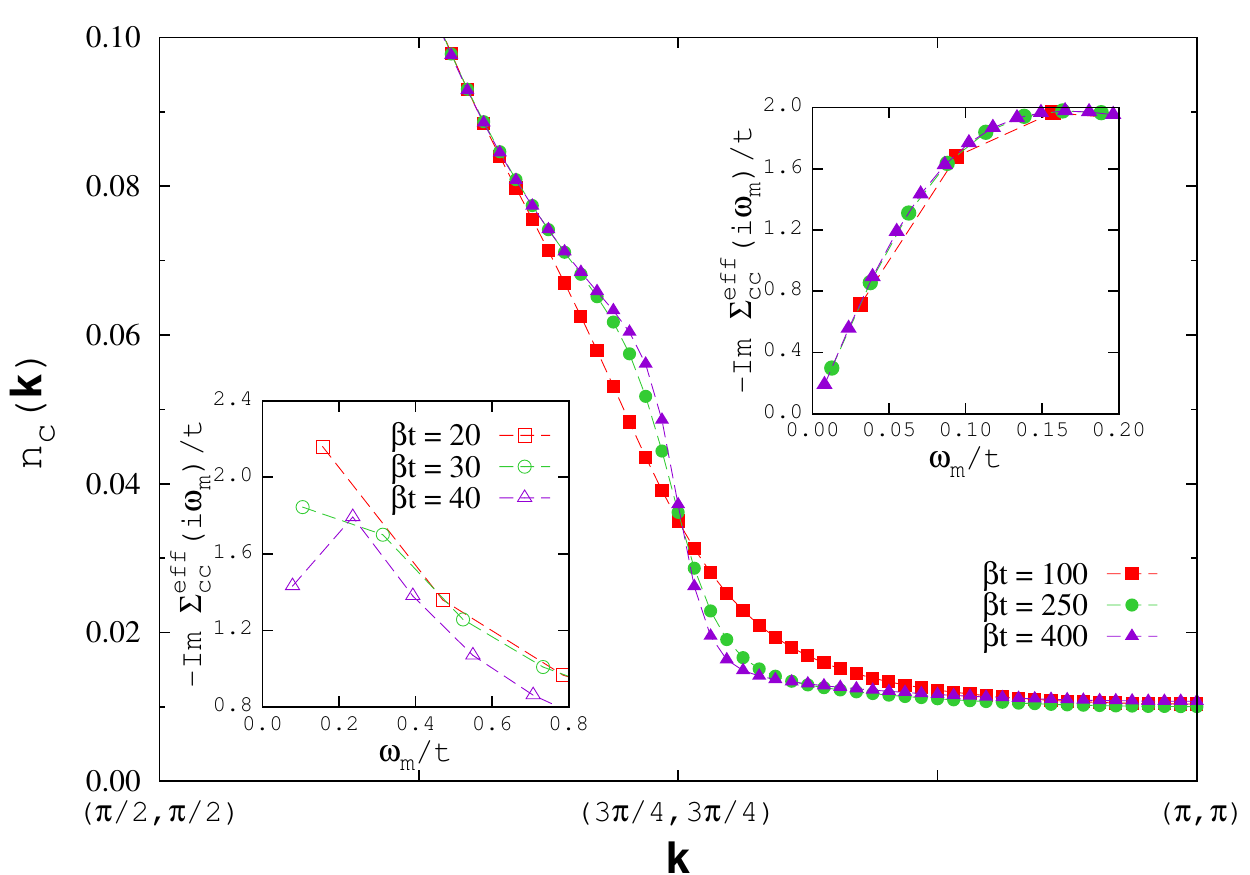}
\end{center}
\caption
{(Color online) Same as in Fig.~\ref{S_ef} but with $\lambda=0.5$ and 
$\omega_0/t=0.5$. }
\label{S_ef_ph}
\end{figure}

We  now  discuss how the previously found features of the
coherent ground state of the PAM are altered by  the 
electron-phonon coupling.  At zero temperature, the  imaginary part of the
self-energy  of Eq.~(\ref{sigma_ph}) reads,
\begin{equation}
\Sigma''_{cc}(\omega)=-\frac{g^2}{2k}\pi\omega_0 \Bigl\{ 
                       N_c^{+} \bigl[1-\Theta(\omega+\omega_0)\bigr]
                     + N_c^{-}     \Theta(\omega-\omega_0) \Bigr\},   
\label{im}
\end{equation}
with $N_c^{\pm}= N_c^{}(\omega\pm\omega_0)$ and $\Theta(\omega)$ being a usual
Heavyside step function. 
Consequently, as shown in Fig. \ref{re_im}, $ \Sigma''_{cc}(\omega)  $ vanishes in a window   $ -\omega_0  < \omega < \omega_0 $. 
Moreover, the hybridization gap of the heavy fermion state, equally leads to
strong suppression of this quantity  just above $\omega > \omega_0$ (see
Fig. \ref{re_im}).  Away from this energy window  the imaginary part of the self-energy is finite thereby producing 
an incoherent high-energy background visible in the single particle spectral function (see Fig. \ref{AN_full_ph}).
A  measure of this incoherent background is obtained from the value of
$n_c({\pmb k}=(\pi,\pi))$ shown in Fig. ~\ref{S_ef_ph}. 
Finally, since the imaginary part of the self energy vanishes at the Fermi energy,
a well define QP is formed  and the corresponding  QP residue $Z$ can be extracted from the real part of self-energy,
\begin{equation}
\Sigma'_{cc}(\omega)=-\frac{1}{\pi} {\mathcal P} \int_{-\infty}^{\infty} 
                 d\omega'\frac{\Sigma''_{cc}(\omega')}{\omega-\omega'}, 
\label{re}
\end{equation}
and with the use of Eq. (\ref{Z}).

\begin{figure}[t!]
\begin{center}
\includegraphics[width=0.45\textwidth]{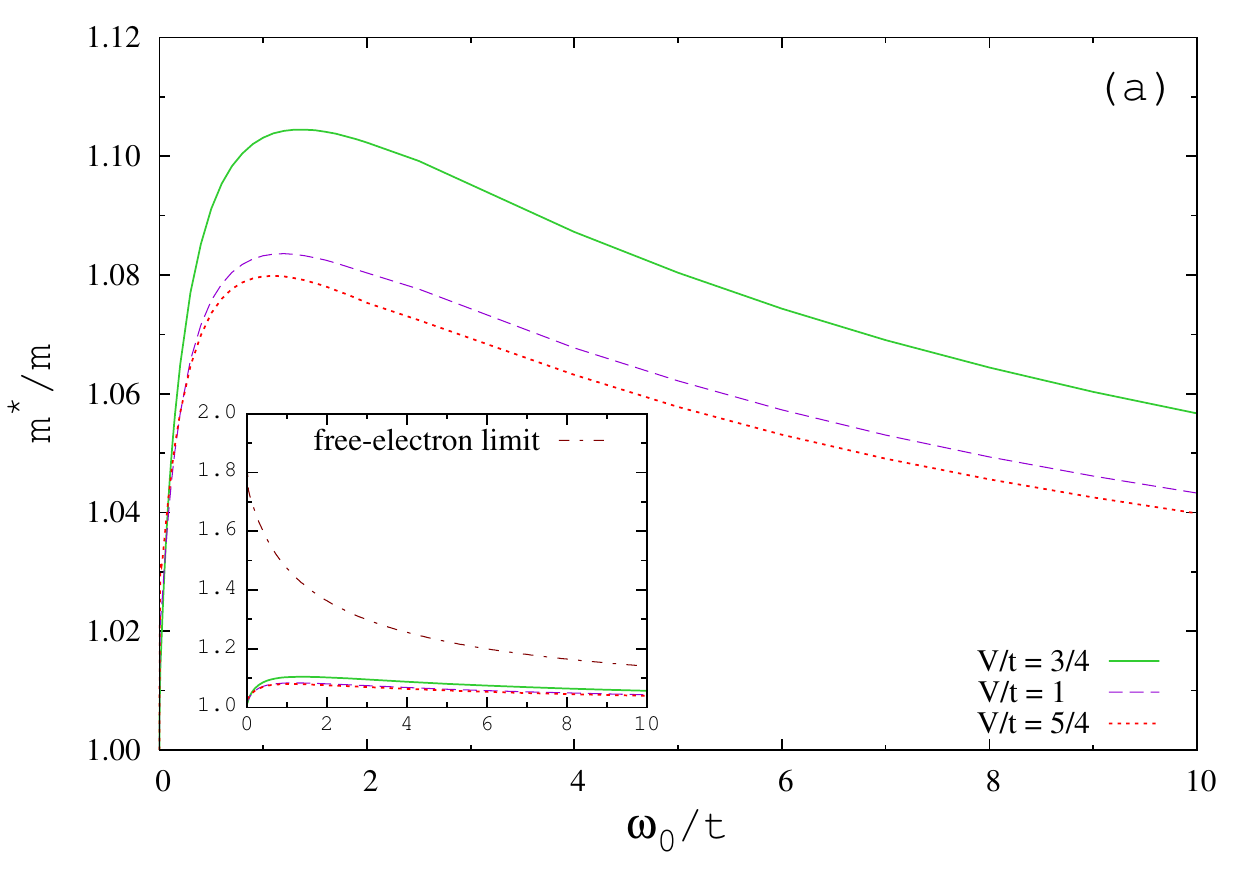}
\includegraphics[width=0.45\textwidth]{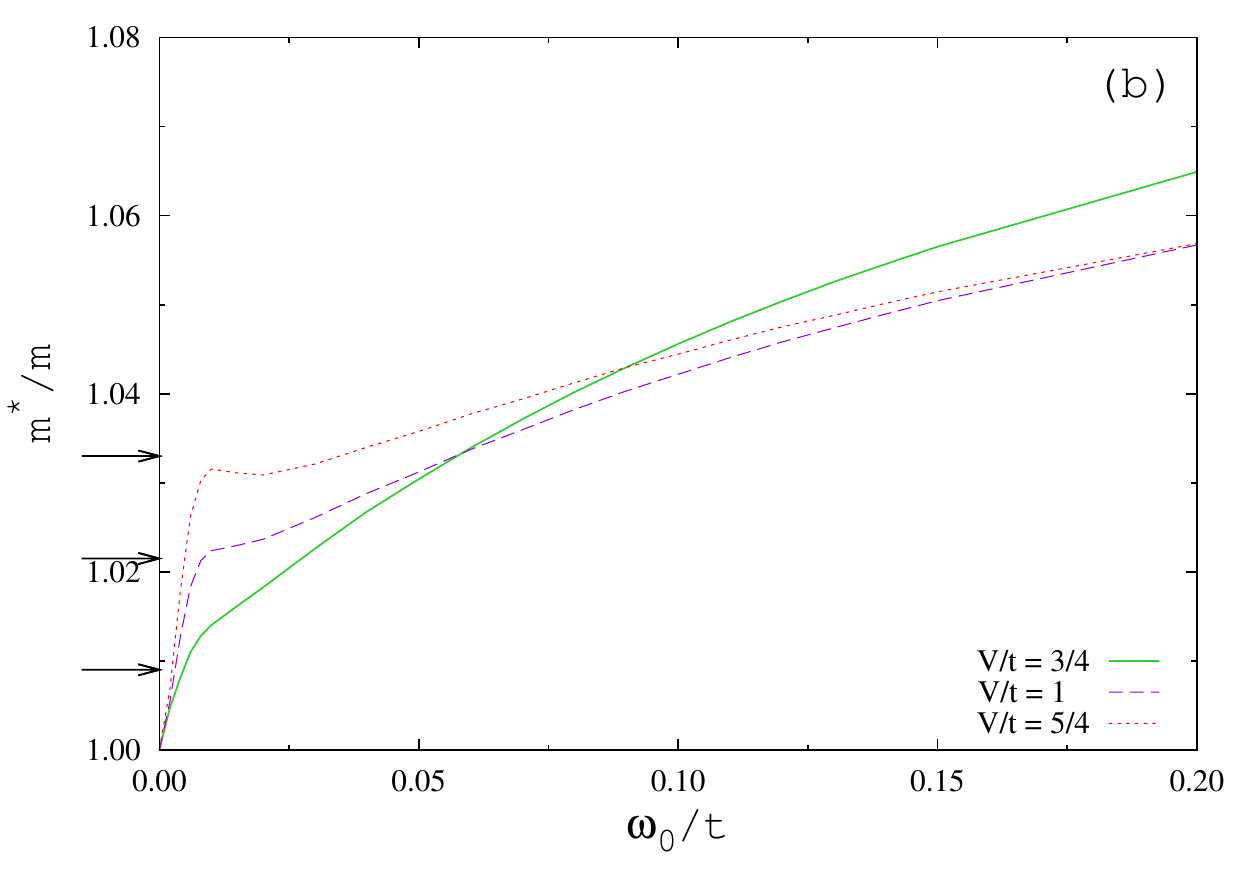}
\end{center}
\caption
{(Color online) Enhancement of the effective mass $m^*$ obtained for a fixed electron-phonon coupling $\lambda=0.5$
  at $\beta t=400$ for representative values of the hybridization $V$. 
  (a) Overall behavior of $m^*$. Inset: enhancement of $m^*$ in the heavy fermion state is marginal as
  compared to the free-electron limit.  
 (b) Low-frequency limit $\omega_0/t\le0.2$. The arrows indicate $m^*$
  estimated from Eq.~(\ref{mstar_intro.eq}). 
}
\label{m_V}
\end{figure}

\begin{figure}[t!]
\begin{center}
\includegraphics[width=0.45\textwidth]{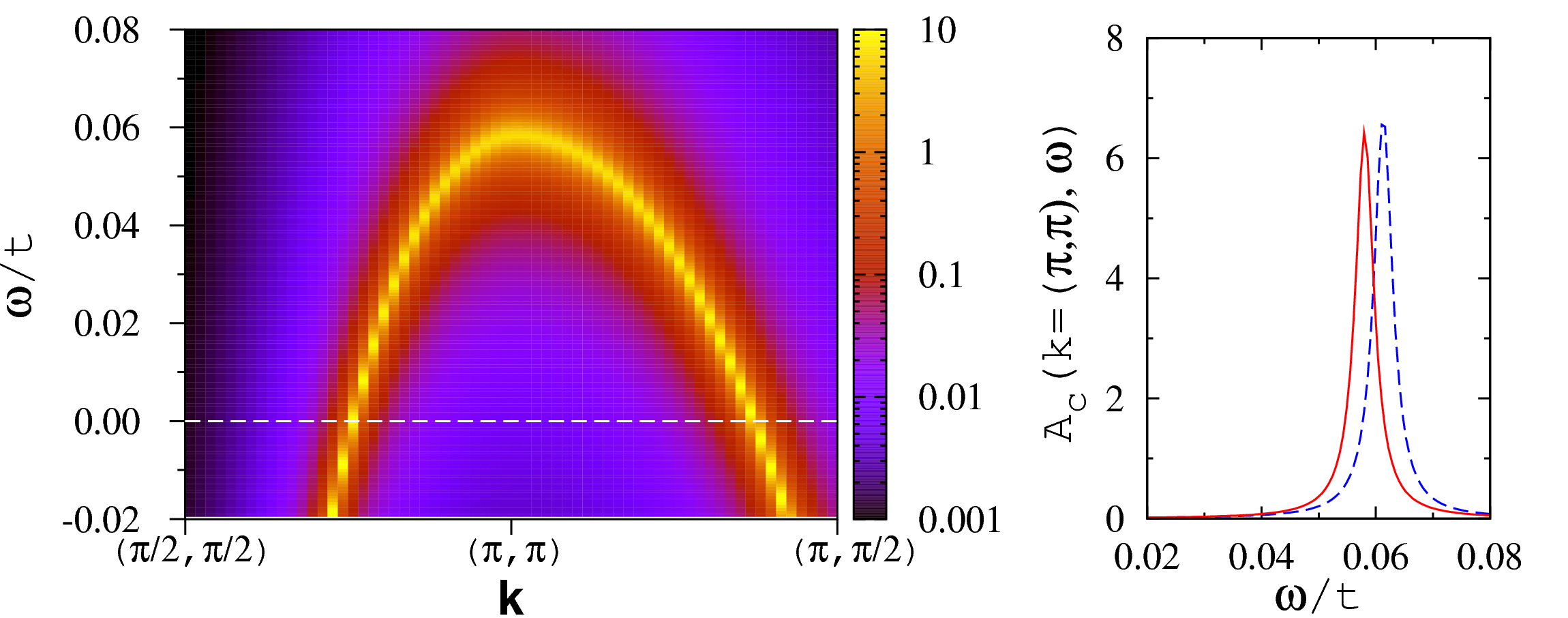}\\
\includegraphics[width=0.45\textwidth]{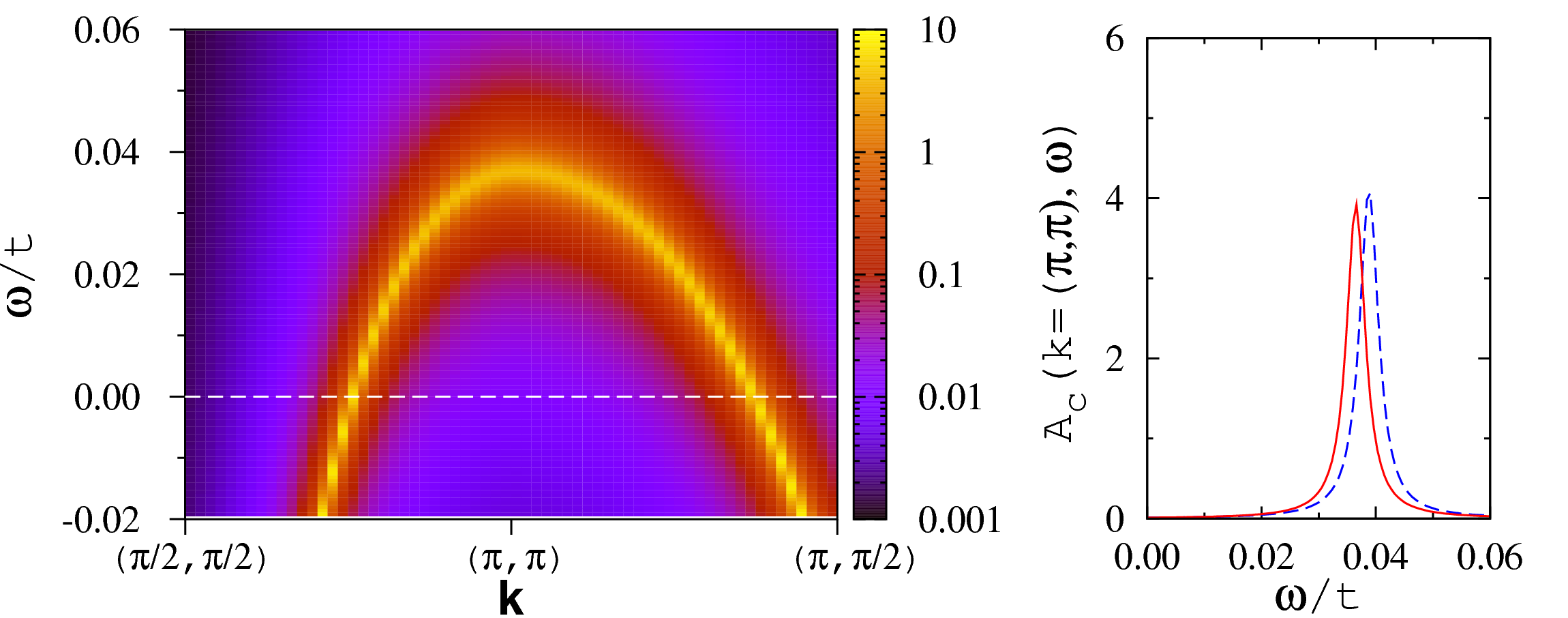}  \\
\includegraphics[width=0.45\textwidth]{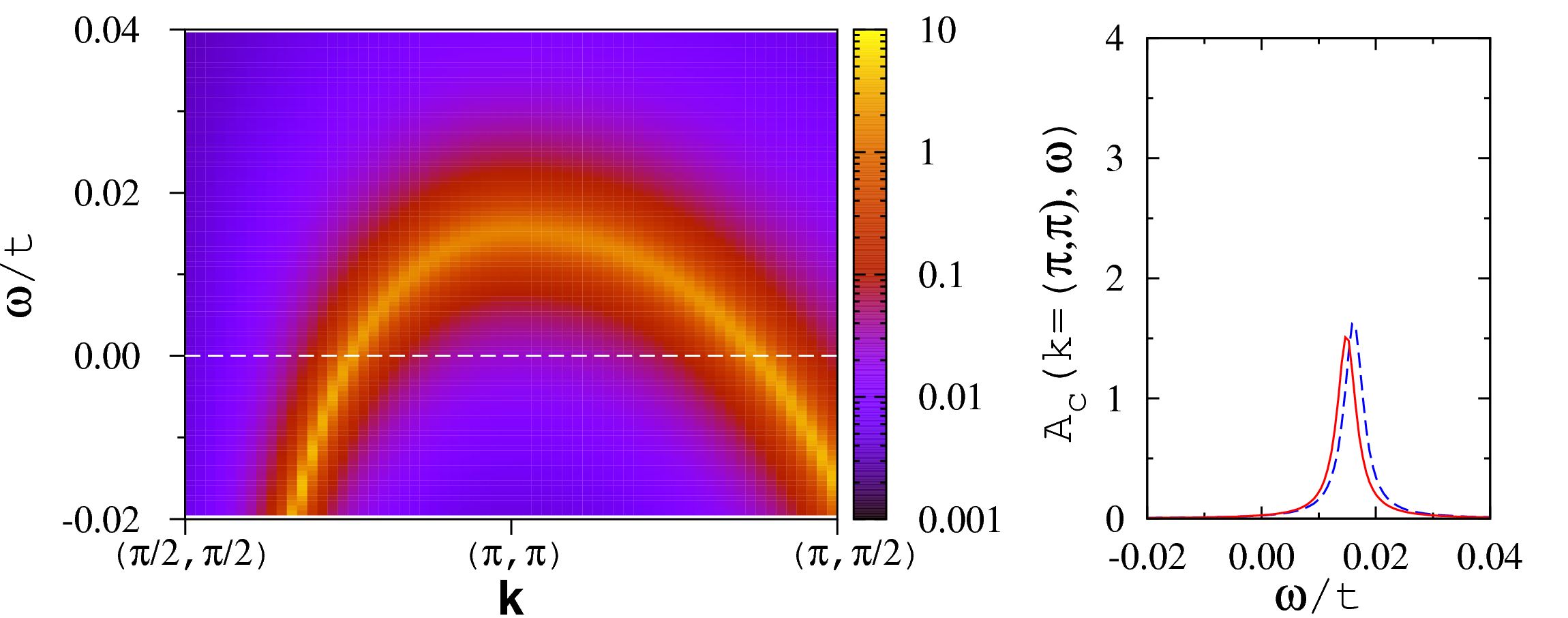}
\end{center}
\caption
{(Color online) Low-energy part of the $c$-electron spectral function $A_c({\bf k},\omega)$ (left)
  and the QP peak at ${\bf k}=(\pi,\pi)$ (right) obtained with $\lambda=0.5$ and $\omega_0/t=0.5$ at $\beta t= 400$: 
$V/t=1.25$ (top), $V/t=1$ (middle), and $V/t=0.75$ (bottom).     
For comparison, the corresponding $A_c({\bf k}=(\pi,\pi),\omega)$  in the
absence of phonons (dashed line) is shown. 
}
\label{ANph}
\end{figure}

We turn now to the most important aspect of our analysis  namely the mass renormalization due to the 
inclusion of the rainbow diagrams. This quantity is plotted in Fig. \ref{m_V} at $\lambda = 0.5 $ and as a function of 
phonon frequency $\omega_0$. Here one finds that in comparison to the free-electron case, shown
as inset in Fig.~\ref{m_V}(a), the overall mass renormalization is extremely small
especially in the adiabatic case. As depicted in Fig. \ref{ANph}, this weak mass renormalization is also
apparent in delicate flattening of the QP band producing very small shift in
the peak position of the QP pole in the vicinity of ${\pmb k}=(\pi,\pi)$.

To understand  the origin of this weak mass renormalization,  we can start from the hybridized band picture  as obtained 
from the SB approach which up  to a constant produces a mean-field Hamiltonian: 
\begin{equation}
	H_{SB} = \sum_{ {\pmb k  }, \sigma, n= \pm }  E_n( {\pmb k} ) \eta^{\dagger}_{{\pmb k},\sigma,n   }
       \eta_{  {\pmb k}, \sigma, n  }   + H_{ph},
\end{equation}
where the QP energies  $E_n( {\pmb k})$ are given by Eq. (\ref{Epm.eq}),
$\eta^{\dagger}_{{\pmb k},\sigma,n} $ are the corresponding QP  operators,
\begin{equation}
\begin{gathered}
	\eta_{ {\pmb k}, \sigma, +} = u_{\pmb k} c_{ {\pmb k}, \sigma} - v_{\pmb k} \tilde{f}_{ {\pmb k}, \sigma},  \\
	\eta_{ {\pmb k}, \sigma, -} = u_{\pmb k} c_{ {\pmb k}, \sigma} + v_{\pmb k} \tilde{f}_{ {\pmb k}, \sigma}, 
\end{gathered}
\end{equation}
while  $u( {\pmb k} )$ and $v( {\pmb k} )$ are the coherence factors: 
\begin{equation}
\begin{gathered}
    u( {\pmb k} ) = \frac{1}{\sqrt{2}} 
      \left(   1 +  \frac{\epsilon( {\pmb k} ) - \tilde{\epsilon}_f } {E({\pmb k})} \right)^{1/2}, \\
    v( {\pmb k} ) = \frac{1}{\sqrt{2}} 
      \left(   1 -  \frac{\epsilon( {\pmb k} ) - \tilde{\epsilon}_f } {E({\pmb k})} \right)^{1/2}, 
\end{gathered}
\end{equation}
where $ E({\pmb k}) = \sqrt{[{\tilde \epsilon_f} - \epsilon({\pmb k})]^2 + 4 (z_{}V)^2 } $.
Introducing the  raising and lowering operators:
\begin{equation}
	a_{\pmb i}^{\dag} =  \frac{\omega_0 M {\hat Q}_{\pmb i} -  i {\hat P}_{\pmb i} } { \sqrt{2
            \omega_0 M }}, 
\qquad 
       a_{\pmb i} =  \frac{\omega_0 M {\hat Q}_{ \pmb i} +  i {\hat P}_{\pmb i} } { \sqrt{2
            \omega_0 M }}, 
\end{equation}
the phonon  part of our Hamiltonian reads: 
\begin{equation}
	H_{ph} =
	g \sqrt{\frac{\omega_0}{2 k } } \sum_{ {\pmb q} } \rho({\pmb q} ) 
     \left(  a^{\dagger}_{\pmb q} + a^{}_{- \pmb q}\right)    + \omega_0 \sum_{\pmb q}  a^{\dagger}_{\pmb q}a^{}_{\pmb q},
\end{equation}
where  the phonons couple to  the charge density, 
\begin{equation}
	\rho({\pmb q}) = \frac{1} {\sqrt{N}}  \sum_{ {\pmb k}, \sigma } c^{\dagger}_{ {\pmb k},\sigma} 
          c^{}_{ {\pmb k} + {\pmb q} ,\sigma }. 
\end{equation}
In the low  phonon frequency limit,  it is appropriate to  rewrite the charge density in terms of the 
heavy quasiparticles, 
\begin{align}
    \rho({\pmb q})  &=  \frac{1}{\sqrt{N}} \sum_{ {\pmb k}, \sigma } \nonumber \\
         \Bigl( & u_{\pmb k} u_{{ \pmb k} + { \pmb q }} \eta^{\dagger}_{ {\pmb k}, \sigma, +} 
                                                \eta^{}_{ {\pmb k} + {\pmb q} , \sigma, +} +  
          v_{\pmb k} v_{{ \pmb k} + { \pmb q }} \eta^{\dagger}_{ {\pmb k}, \sigma, -} 
                                                \eta^{}_{ {\pmb k} + {\pmb q} ,
                                                  \sigma, -}   \nonumber \\ 
         + &u_{\pmb k} v_{{ \pmb k} + { \pmb q }} \eta^{\dagger}_{ {\pmb k}, \sigma, +} 
                                                \eta^{}_{ {\pmb k} + {\pmb q} , \sigma, -} + 
          v_{\pmb k} u_{{ \pmb k} + { \pmb q }} \eta^{\dagger}_{ {\pmb k}, \sigma, -} 
                                                \eta^{}_{ {\pmb k} + {\pmb q} ,
                                                  \sigma, +} \Bigr),  
\end{align}  
where intra and interband transitions are apparent. For $\omega_0 < T_{\rm coh} $  we can retain only the intra band 
transitions within the lower hybridized band,  $ E_{-}({\pmb k}) $, thus obtaining an effective low energy Hamiltonian: 
\begin{align}
	H_{eff} &= \sum_{ {\pmb k  }, \sigma }  E_{-}( {\pmb k} ) \eta^{\dagger}_{ {\pmb k}, \sigma, - }
       \eta^{}_{ {\pmb k}, \sigma, -  }   +  \omega_0 \sum_{\pmb q}  a^{\dagger}_{\pmb q}a^{}_{\pmb q} \nonumber \\
        &+  g \sqrt{\frac{\omega_0}{2 k } } \frac{1}{\sqrt{N}} \sum_{ {\pmb k} {\pmb q}, \sigma } 
          v_{\pmb k} v_{{ \pmb k} + { \pmb q }} \eta^{\dagger}_{ {\pmb k}, \sigma, -} 
                                                \eta^{}_{ {\pmb k} + {\pmb q} , \sigma, -}   
     \left(  a^{\dagger}_{\pmb q} + a^{}_{- \pmb q} \right).   
\end{align}
This corresponds to a single band problem  the band width being  $WZ$ where $W$ is the bare band width, and with a
renormalized electron-phonon coupling $ g \rightarrow g Z $.  Here, we have
used the  fact that in the adiabatic limit momentum transfer ${\pmb q}\to 0$ and  $Z = | v( { \pmb k} ) |^2 $. 
Hence, we arrive at Eq. (\ref{mstar_intro.eq})  discussed in the Sec. \ref{sec:0}.

Let us now focus on the low-frequency limit where all the high-energy interband channels
contributing to QP dressing are frozen due to the hybridization gap [see Fig.~\ref{m_V}(b)]. 
The {\it sudden} downturn in the data is  a temperature effect. Making abstraction of this downturn by first 
taking the limit $ T \rightarrow 0 $ and then  $ \omega_0 \rightarrow 0$, gives
values of $m^*/m$ which as indicated in Fig. \ref{m_V}(b) track the scale $1 +
\lambda Z$  thus confirming the above  argument. 
As the phonon frequency grows beyond the coherence temperature, the single low energy band picture fails and 
interband transitions become progressively important.  Indeed, owing to the smallest
hybridization gap ({\it cf.} inset in Fig.~\ref{re_im}), this effect appears first
in the weakly hybridized regime with $V/t=3/4$.   

Furthermore, as  discussed in Sec. \ref{sec:0},  we can acquire insight in the
high frequency limit, by first taking into account  phonon degrees of freedom and 
then the magnetic impurities, to arrive at the conclusion that also in this limit  the phonons have very little 
effect on the scales of the heavy fermion metallic state.  This is again confirmed by Fig.~\ref{m_V}(a)   since 
the overall scale of the plot remains very small in comparison to obtained in the absence of  magnetic impurities.   
In fact, we have checked by varying $0\leq\omega_0/t\leq 10$ and
$0\leq\lambda\leq 1$ that the overall variation of the $Z$ factor remains very small. Consequently, we conclude that  the scales of the 
 heavy fermion metallic state are {\it  protected} within the ME approximation  from the electron-phonon interaction both
in the adiabatic and antiadiabatic  limits.   This  sets the stage for the importance  of vertex corrections 
which is the subject of Sec. \ref{sec:2}.

\section{\label{sec:2} Dynamical mean field approximation}   

We use a recently developed  generalization of the weak coupling 
CT-QMC algorithm   to include phonon degrees of  freedom as impurity solver for the DMFT. \cite{Rubtsov05,Assaad07}   
This algorithm relies on integrating out the phonon degrees of freedom at the expense of a 
retarded attractive density-density interaction in which one expands. This approach to include phonons in
model Hamiltonians has been tested extensively in  the framework of the  one-dimensional Holstein model. \cite{Assaad08} 
We refer the reader to Ref. \onlinecite{Assaad07} for a detailed discussion of the algorithm.  
We have used a stochastic analytical continuation scheme to carry  the rotation from imaginary to real times. 
\cite{Beach04a}

\begin{figure}[t!] 
\begin{center}
\includegraphics*[width=0.45\textwidth]{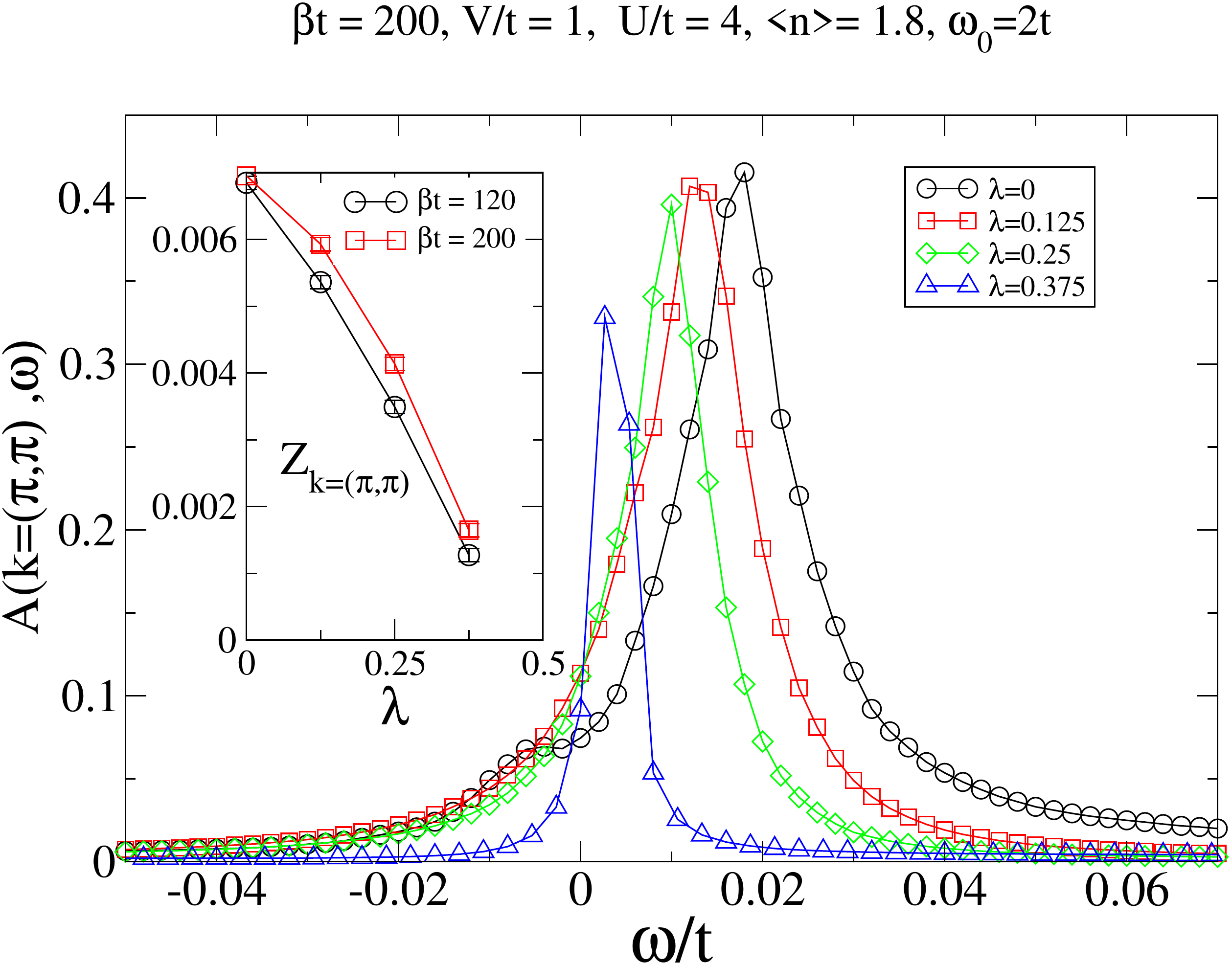}
\end{center}
\caption{(Color online) $c$-electron spectral function $A_c({\pmb k}=(\pi,\pi) ,\omega)$  as a 
function of the electron-phonon coupling.   Inset: QP residue as obtained from fitting the 
$ \beta t = 120 $  ($\beta t = 200 $) data to the form of Eq. (\ref{QMC_fit.eq}) in the range   $  30 < \tau t < 50 $ 
($  30 < \tau t < 70 $), respectively.   }   
\label{QMC_pp_w_scan.fig}  
\end{figure}

We begin with Fig. \ref{QMC_pp_w_scan.fig} which plots the conduction electron
single-particle spectral function at our lowest temperature, 
$\beta t = 200$,  and at the ${\pmb k} = (\pi,\pi)$ point.  As argued in the previous section  the position of the peak as well as 
its residue  is a  measure of the coherence temperature $T_{\rm coh}$. One can estimate the QP  residue directly from the 
low temperature imaginary time Green's function by noting that  
\begin{equation}
	  G({\pmb k},\tau)  \rightarrow   - e^{ - \tau  \Delta_{qp}({\pmb k} ) }   
   | \langle \Psi_0^{N+1} | c^{\dagger}_{ {\pmb k},\sigma} | \Psi_0^N \rangle|^2 
  \text{ for } \tau t >> 1. 
\label{QMC_fit.eq}
\end{equation}
Here, $\Delta_{qp}({\pmb k} )  =  E_0^{ N+1 } ( \pmb k ) - E_0^{N+1} $ where  $ E_0^{ N+1 }  ( {\pmb k }) $ 
corresponds to the energy eigenstate with particle number  $N+1$ and momentum 
$ {\pmb k} $ and $ E_0^{N+1} $ is the ground state energy in the 
Hilbert space  with $N$ particles.
The corresponding QP residue extracted by
fitting the imaginary time data to the above form and in a limited range  
$ 1 <<  \tau t  < \beta t /2 $ is shown in the inset in Fig. \ref{QMC_pp_w_scan.fig}. 
The fact that we obtain   approximatively the same results for $\beta t = 120
$ and $\beta t = 200$ confirms that both the temperatures lie below $T_{\rm coh}$ of this heavy fermion state.   

\begin{figure}[b!] 
\begin{center}
\includegraphics*[width=0.45\textwidth]{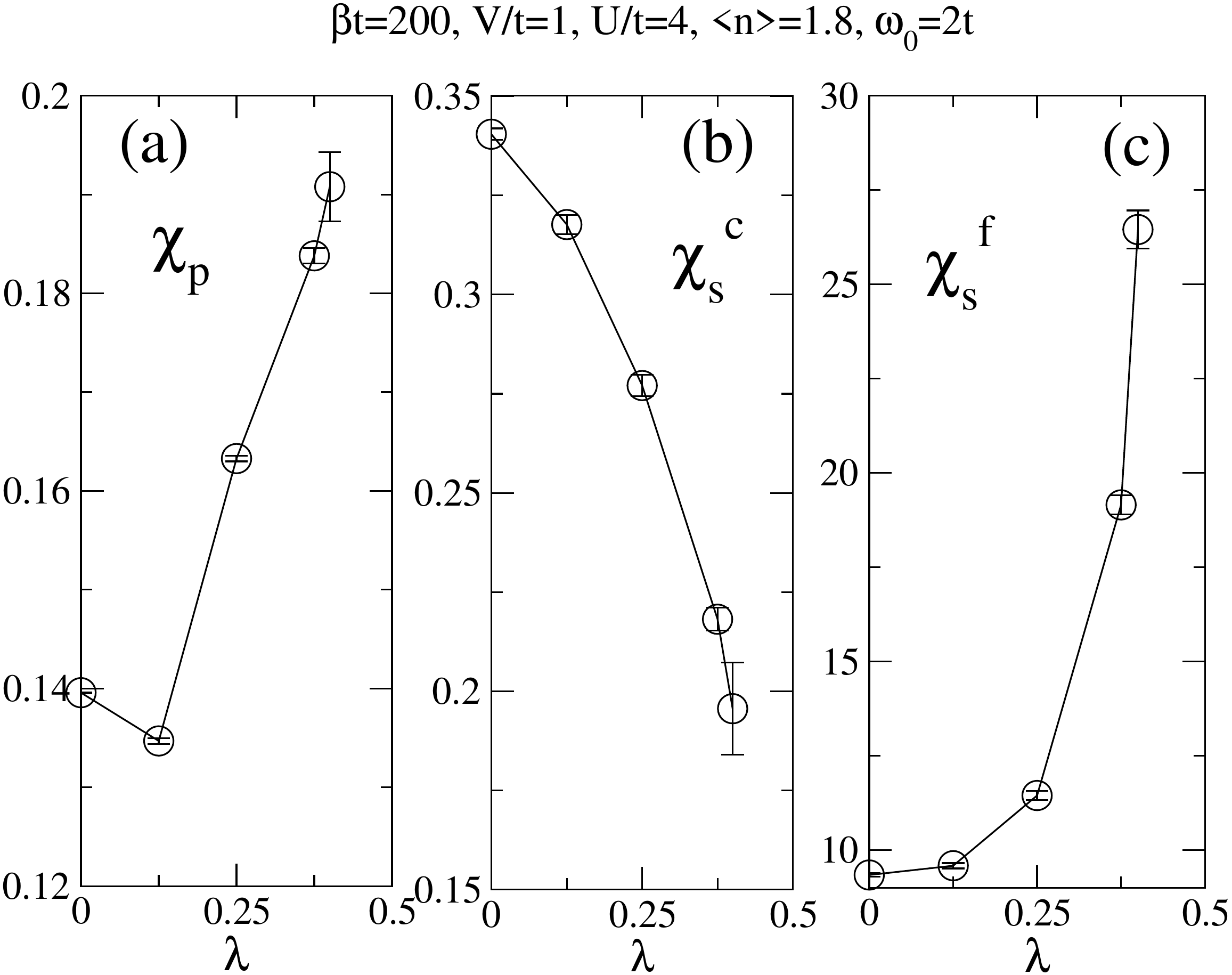}
\end{center}
\caption{  (a) Pair  (b) conduction-  and (c) $f$-spin susceptibilities as  a function of the electron phonon
coupling at our lowest temperature, $\beta t = 200$.  } 
\label{Suscep_g_scan.fig}  
\end{figure} 

As expected, Fig. \ref{QMC_pp_w_scan.fig}  tracks the  evolution of $T_{\rm
  coh}$  --  as defined by the peak position in $A_{c}({\pmb k} = (\pi,\pi), \omega) $ and 
the QP  residue $Z_{ {\pmb k} = (\pi,\pi)}$ -- as a function of growing electron phonon couplings and at fixed 
phonon frequency $\omega_0 = 2t$.  
In the considered  electron-phonon coupling range we observe a considerable decrease of this quantity, 
and the data is consistent with a vanishing $T_{\rm coh}$ at
$\lambda \simeq 0.5$. The fact that we cannot account for this large suppression of $T_{\rm coh}$ within the ME 
approximation highlights the importance of vertex corrections in this parameter range.
\begin{figure}[t!] 
\begin{center}
\includegraphics*[width=0.35\textwidth]{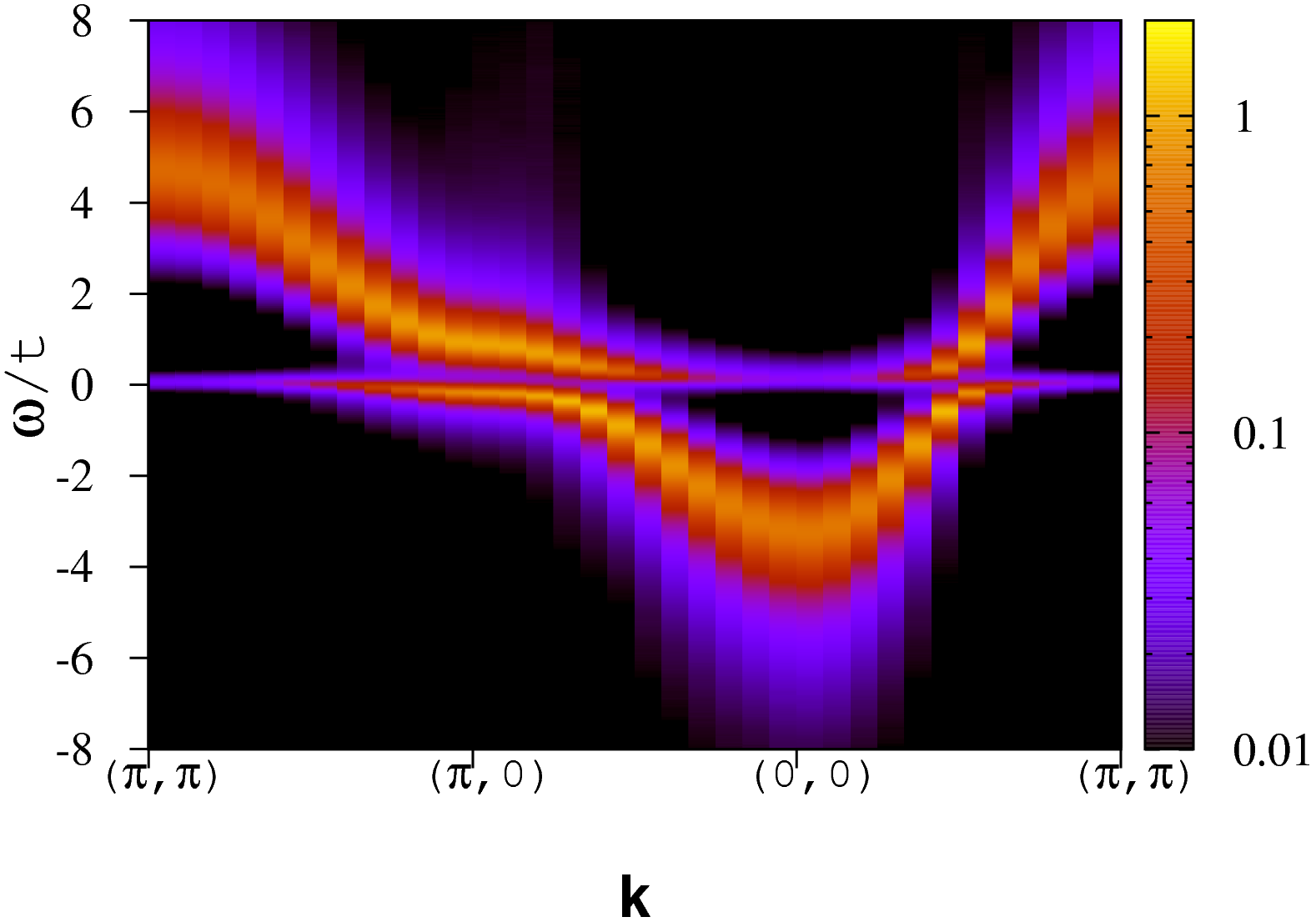} \\
\includegraphics*[width=0.35\textwidth]{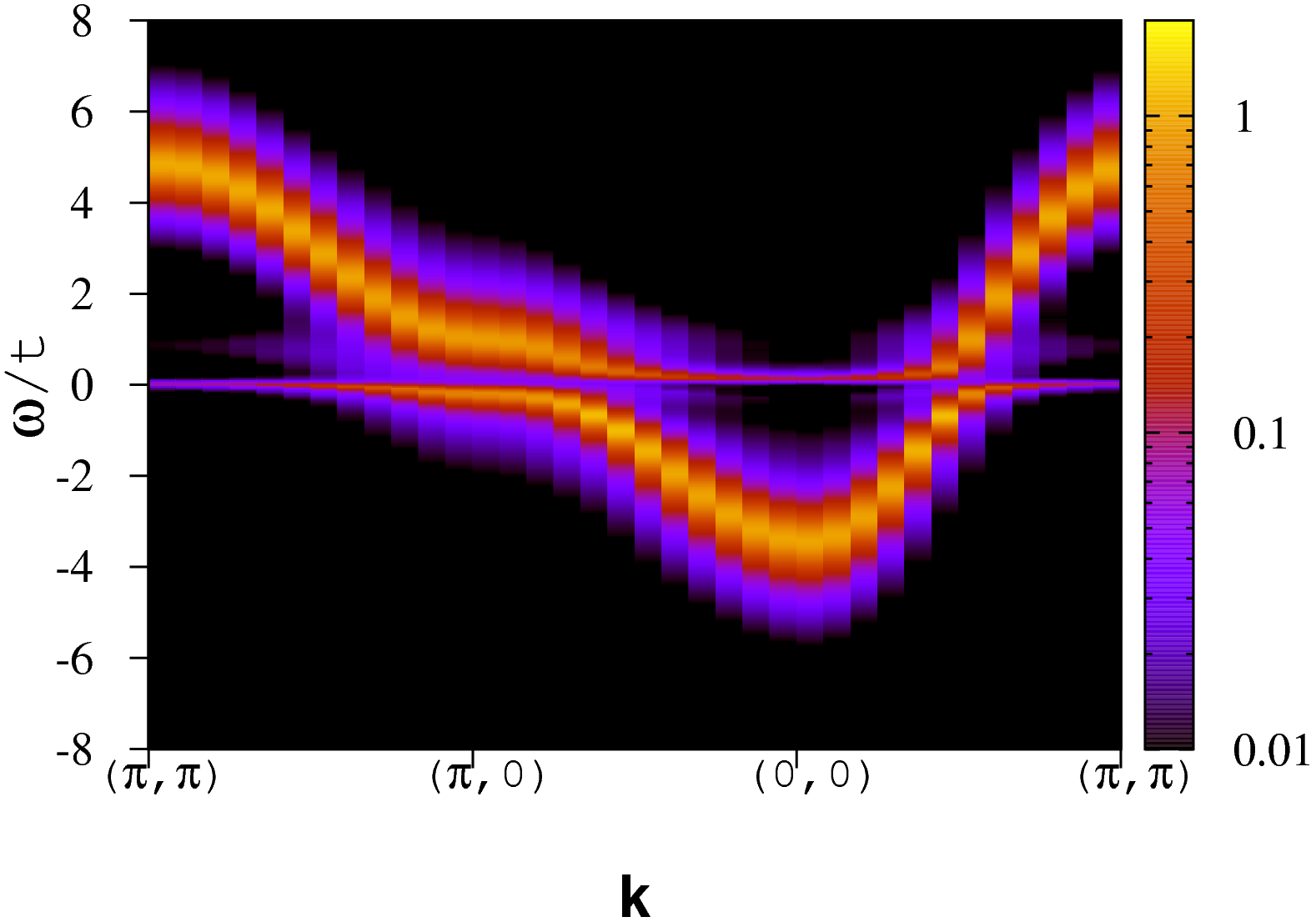} \\
\includegraphics*[width=0.35\textwidth]{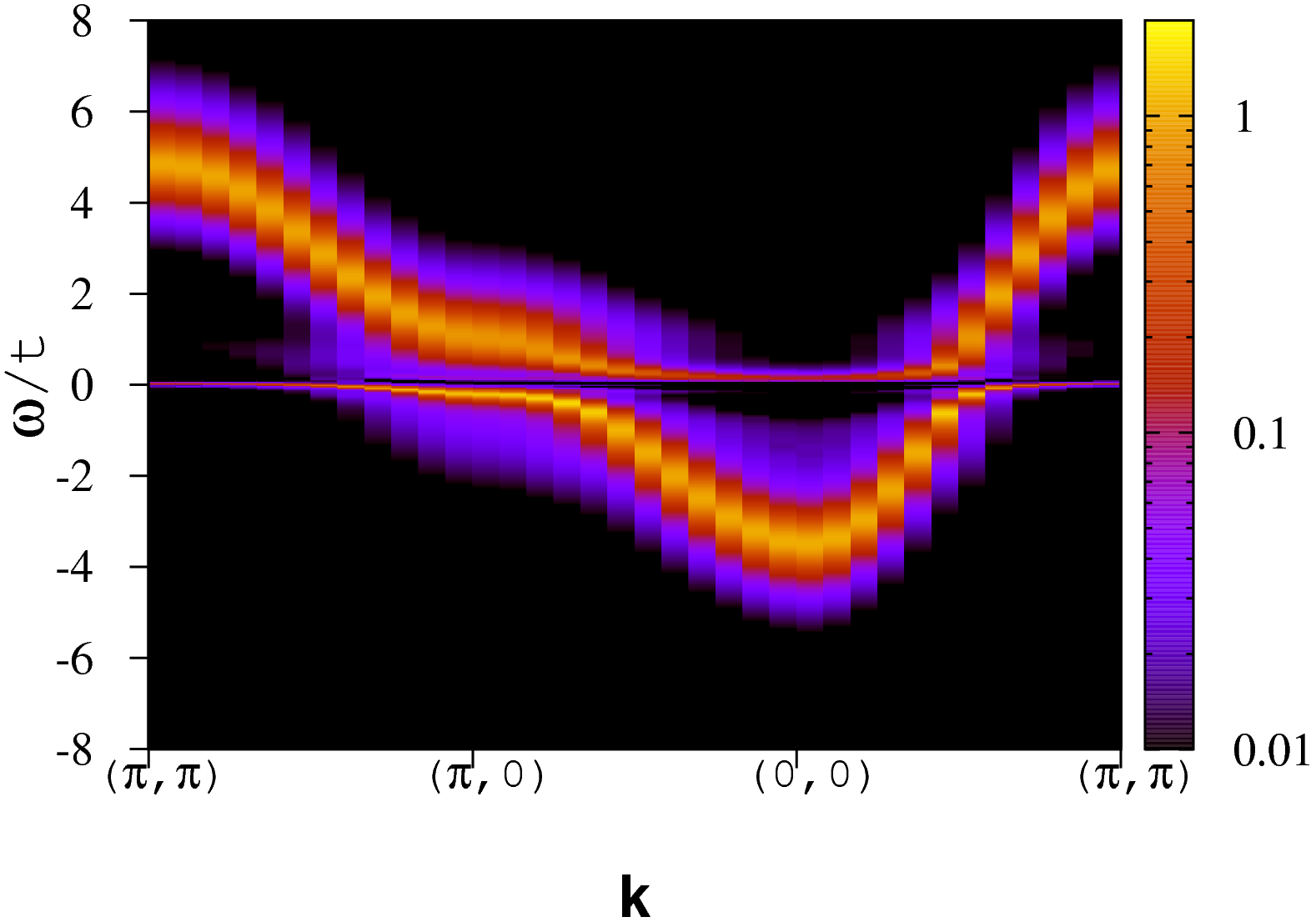} 
\end{center}
\caption{(Color online) $c$-electron single-particle spectral function  in the absence of  phonons. The models parameters 
are given by $V/t=1$, $U/t=4$, $\epsilon_f = \mu$, and $\langle n \rangle = 1.8$. The temperature from top to bottom corresponds 
to $\beta t = 10, 20, \text{and } 120$. 
}
\label{Akom.fig}
\end{figure} 
\begin{figure}[t!]
\begin{center}
\includegraphics*[width=0.35\textwidth]{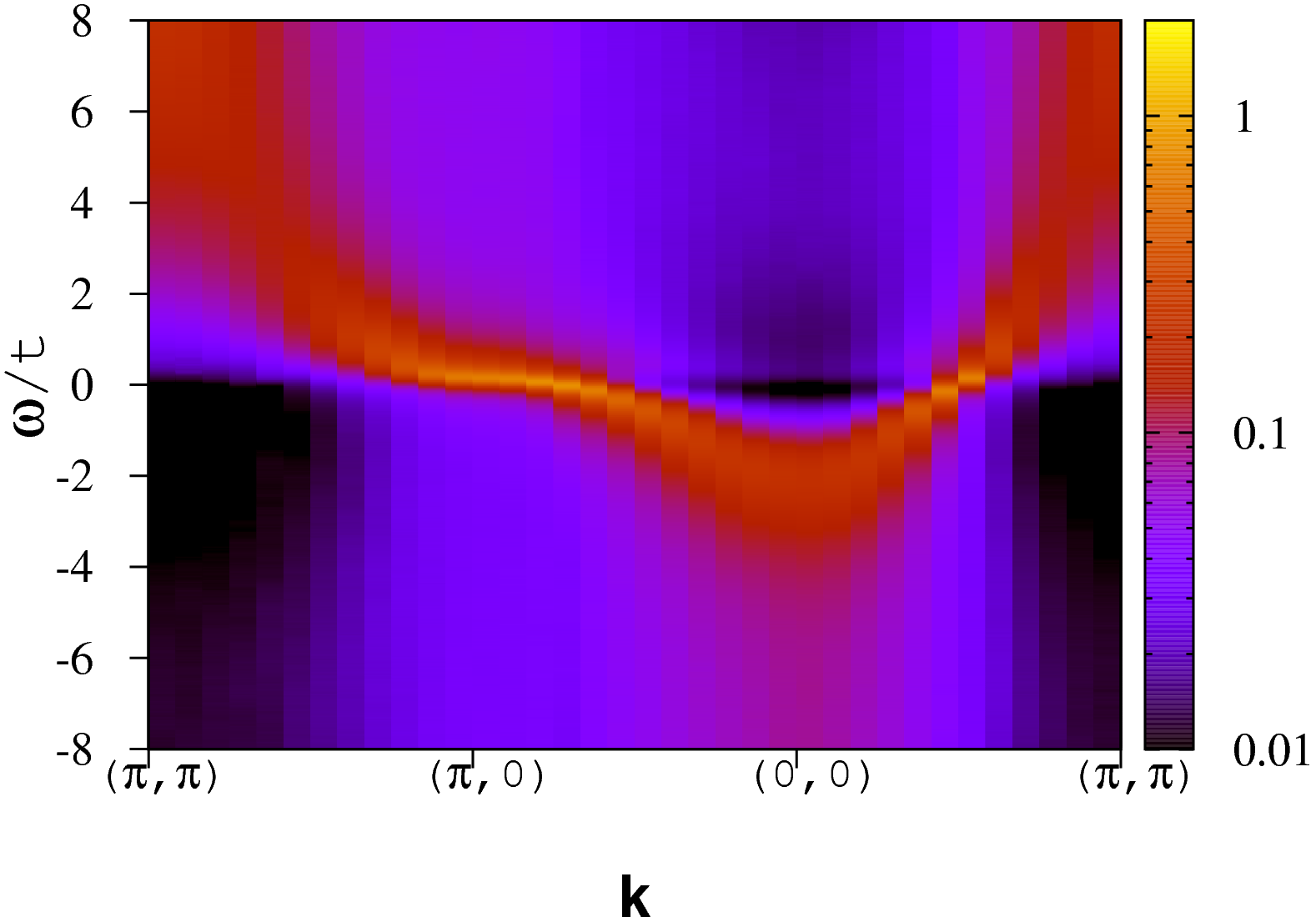} \\
\includegraphics*[width=0.35\textwidth]{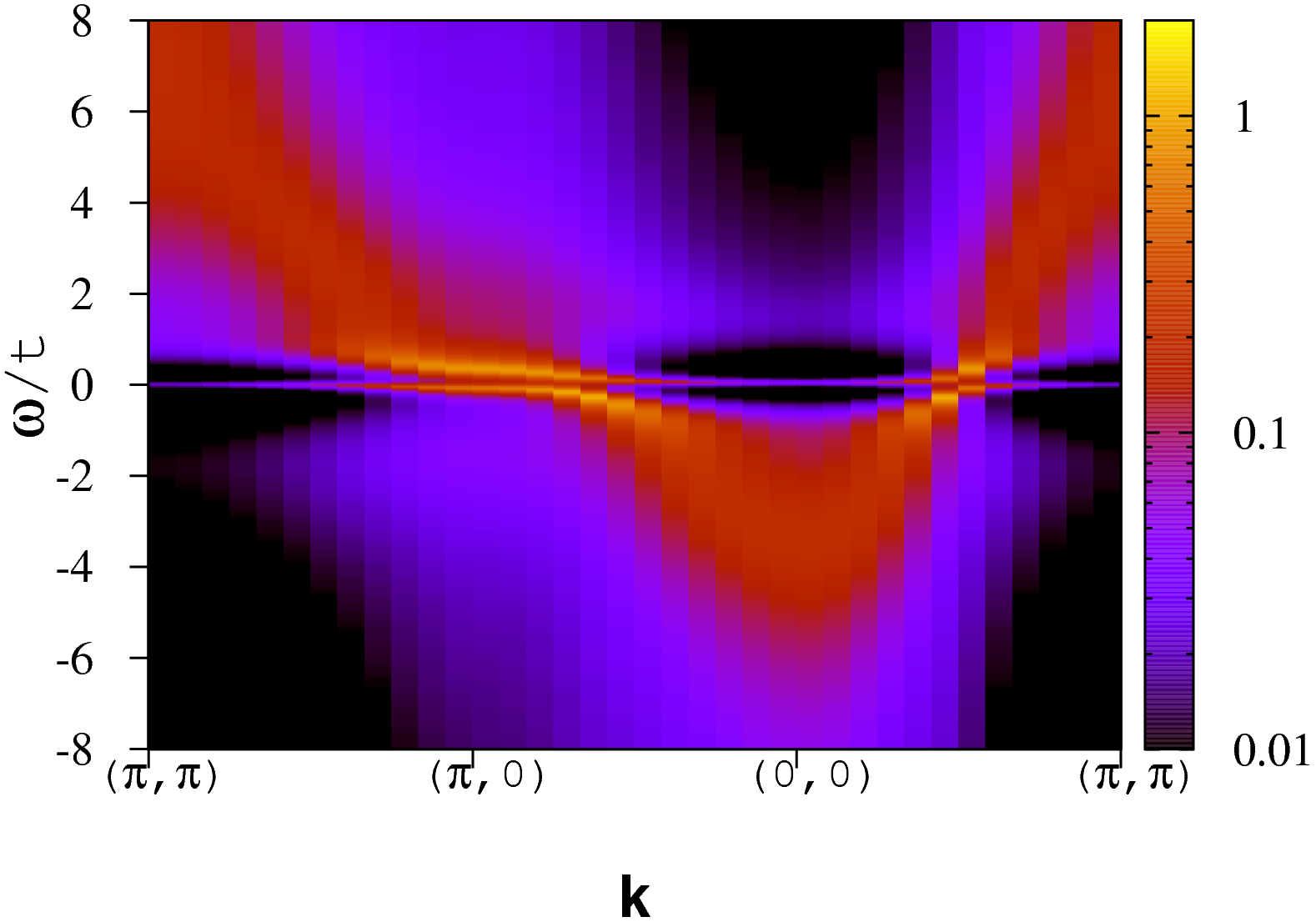} \\
\includegraphics*[width=0.35\textwidth]{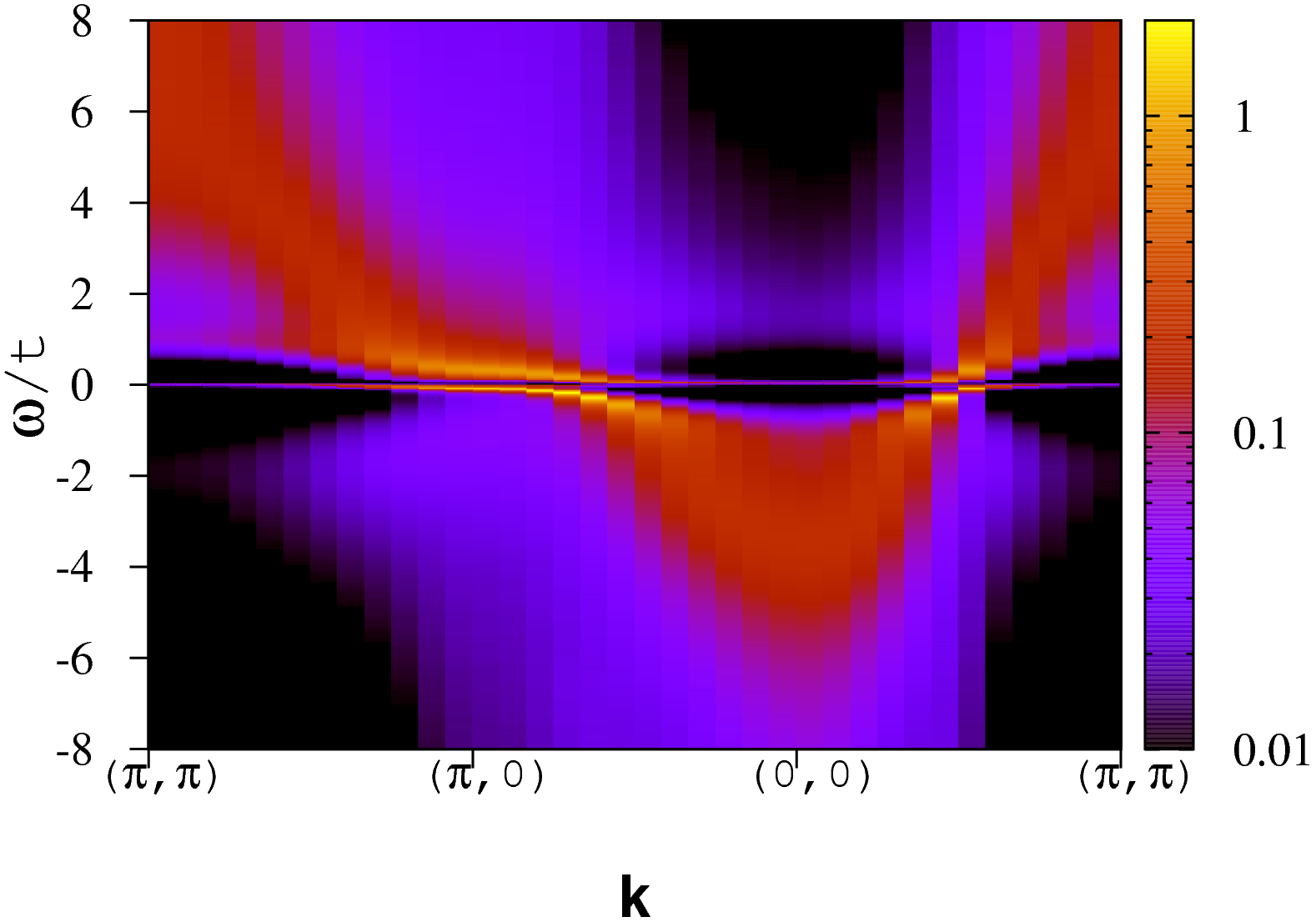}
\end{center}
\caption{(Color online) Same as in Fig.~\ref{Akom.fig} but in the presence of
  phonons with $\lambda = 0.375$ and $\omega_0 = 2t$. }
\label{Akom_el_ph.fig}
\end{figure}
To  confirm  that  the drop in $T_{\rm coh}$ is driven by  electron binding we  have computed the 
local pairing,  $\chi_p$,  and spin susceptibilities, $\chi_s^{\alpha}$  on the cluster as defined by
\begin{equation}
\begin{aligned}
     \chi_p  & =     \int_{0}^{\beta} {\rm d} \tau \langle  
               c^{\dagger}_{\uparrow} (\tau) c^{\dagger}_{\downarrow}(\tau)
               c_{\downarrow}  c_{\uparrow}  \rangle, \\
     \chi_s^\alpha & = \int_{0}^{\beta} {\rm d} \tau \langle  S_z^{\alpha}(\tau)  S_z^{\alpha}  \rangle, 
\label{Local_suscep.eq}
\end{aligned}
\end{equation}
where $\alpha$ stands for the conduction or $f$-electron  in the corresponding impurity problem.   
Those quantities are plotted in Fig. \ref{Suscep_g_scan.fig}. As shown in
Fig. \ref{Suscep_g_scan.fig}(a), after an initial drop, the pair  susceptibility  grows  and  tracks the decrease 
of the coherence temperature.   The growth stems from conduction  electron binding  which originates from the attractive 
retarded interaction  between conduction electrons  mediated by a phonon exchange.  
Regarding a small drop in $\chi_p$ at weak electron-phonon coupling, we  understand this 
drop in terms of  quasiparticle  formation: the bare electron  which enters our definition of the singlet $s$-wave  pair  
acquires a smaller overlap with the QP  as a function of growing $\lambda$. This dressing of the bare electron 
reduces the pair susceptibility and initially overcomes the growth of this quantity due to electron binding.

Electron binding  into $s$-wave singlets as suggested by Fig. \ref{Suscep_g_scan.fig}(a)  lead to a drop of the conduction spin 
susceptibility. This is clearly seen in Fig. \ref{Suscep_g_scan.fig}(b).  Indeed, pairing of conduction electrons   
competes with the Kondo effect in which  conduction electrons form an entangled state with the $f$-electrons thereby 
screening the magnetic impurities.  Therefore, growing values of $\lambda$ are linked to a breakdown of the Kondo singlets  and 
hence a growth of the $f$-spin susceptibility as shown in
Fig. \ref{Suscep_g_scan.fig}(c). This behavior should be contrasted with results obtained in complementary
studies of the PAM extended by phonon coupling to the $f$-electrons.\cite{Ono05}  In this
case, electron-phonon coupling generates not only pairing between $f$-orbitals
but also indirectly via finite mixed valence induces binding of the conduction
electrons.

Furthermore, Figs. \ref{Akom.fig}  and \ref{Akom_el_ph.fig} compare the 
temperature dependence of the  $c$-electron single-particle spectral functions  with and without phonon degrees of freedom. 
In the absence of  the electron phonon interaction, it has been shown \cite{Beach08} that  signatures 
of the hybridized bands in the single-particle spectral function emerge  below the single-ion Kondo temperature. 
On the other hand, it is only below the coherence temperature that a well define hybridization  gap is visible. 
As shown in  Ref. \onlinecite{Beach08},  this hybridization gap leads to a plateau feature in the magnetization curve.   
The plots in  Fig. \ref{Akom.fig}  do not include  coupling to the  phonon degrees of freedom 
and point to the gradual formation of the hybridization gap.  It is only at  the lowest shown temperature, 
$\beta t = 120$, that a clear hybridization gap is apparent.  The SB mean-field analysis 
gives  $T_K /t \simeq 0.7$  for this parameter set.  Hence, for  the temperature  range  considered in Fig. \ref{Akom.fig}
and in the absence of electron-phonon interaction, we always expect signatures of the  hybridized band structure when  
the data is taken below the Kondo temperature.

\begin{figure}[t!]
\begin{center}
\includegraphics*[width=0.45\textwidth]{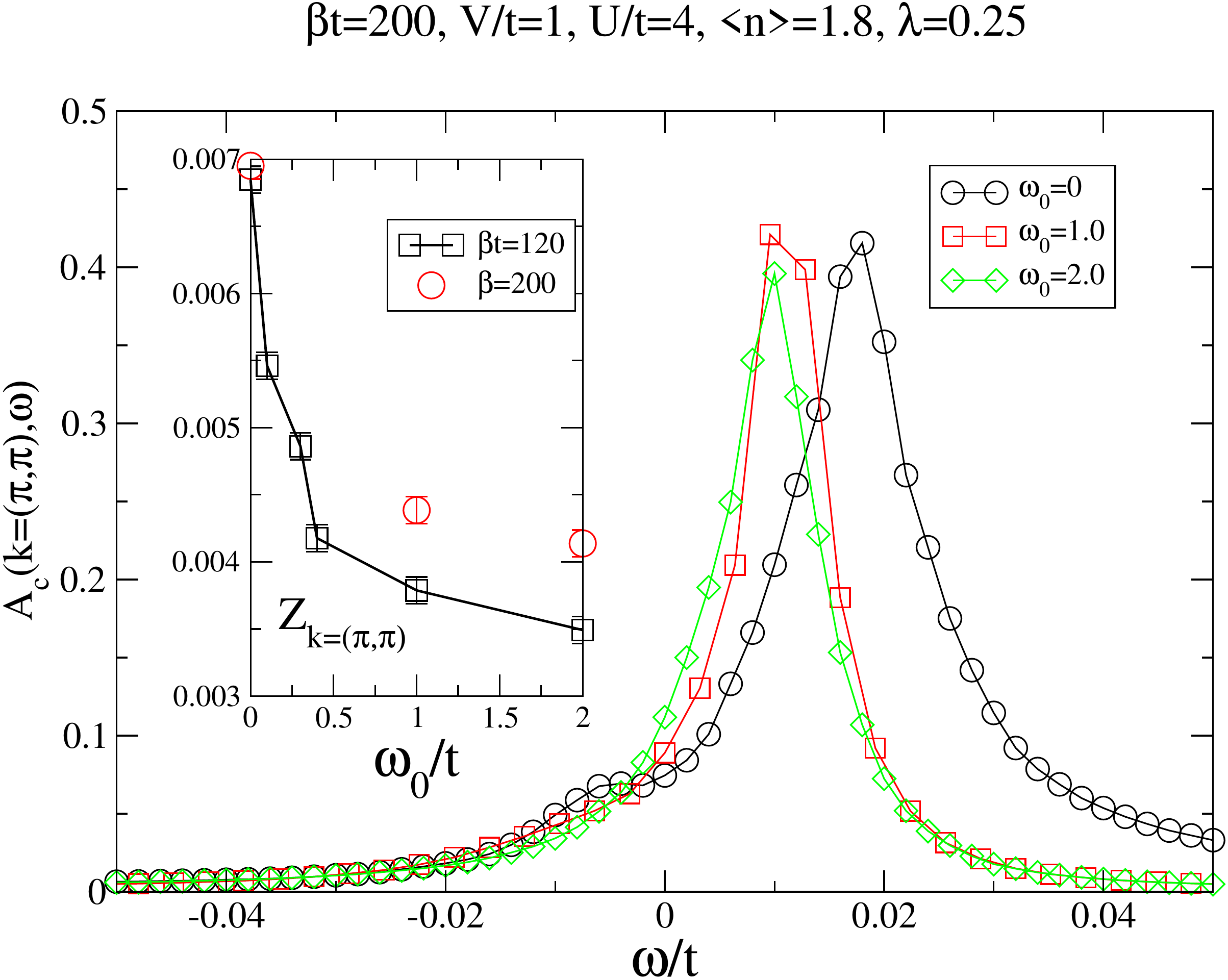}
\end{center}
\caption{(Color online)  
Single-particle spectral function of the conduction electrons  and QP residue (inset)  at fixed electron-phonon coupling, 
$\lambda = 0.25$, and as a function of  phonon frequency. The data points at
$\omega_0 = 0 $  correspond to the case without phonons.  
}
\label{QMC_pp_w_omscan.fig}
\end{figure}

Turning on  the electron phonon interaction to $\lambda = 0.375$   and setting the phonon frequency to $\omega_0 = 2t$ 
reduces $T_{\rm coh}$ -- as measured by the  QP residue  (see Fig. \ref{QMC_pp_w_scan.fig}) -- by 
roughly an order of magnitude.   Since one expects the Kondo temperature to track the coherence temperature,  a similar 
reduction  of the former quantity is foreseen.  Our highest temperature  spectral function, $\beta t = 10$,  in Fig. 
\ref{Akom_el_ph.fig}  shows no sign of  the hybridization bands thus confirming  the substantial reduction of this 
scale  upon inclusion of the electron-phonon coupling. At this temperature,
only high energy features and an enhancement of the effective mass by the phonon degrees of freedom  is apparent in 
the data.   The Fermi surface of this state accounts \emph{solely} for the
conduction electrons.  Upon lowering the temperature,  a hybridization gap  is
restored in the QP band but its width is substantially reduced as
compared to the corresponding one in Fig.~\ref{Akom.fig}.  Only at our lowest
considered temperature  one can perceive the coherent heavy fermion metallic state  with very \emph{faint} QP peak crossing the Fermi energy in 
the vicinity of the Brillouin  zone corner ${\pmb k} = (\pi,\pi)$ thus
indicating reconstruction of a \emph{large} Fermi surface including both the conduction and $f$-electrons.

\begin{figure}[t!] 
\begin{center}
\includegraphics*[width=0.45\textwidth]{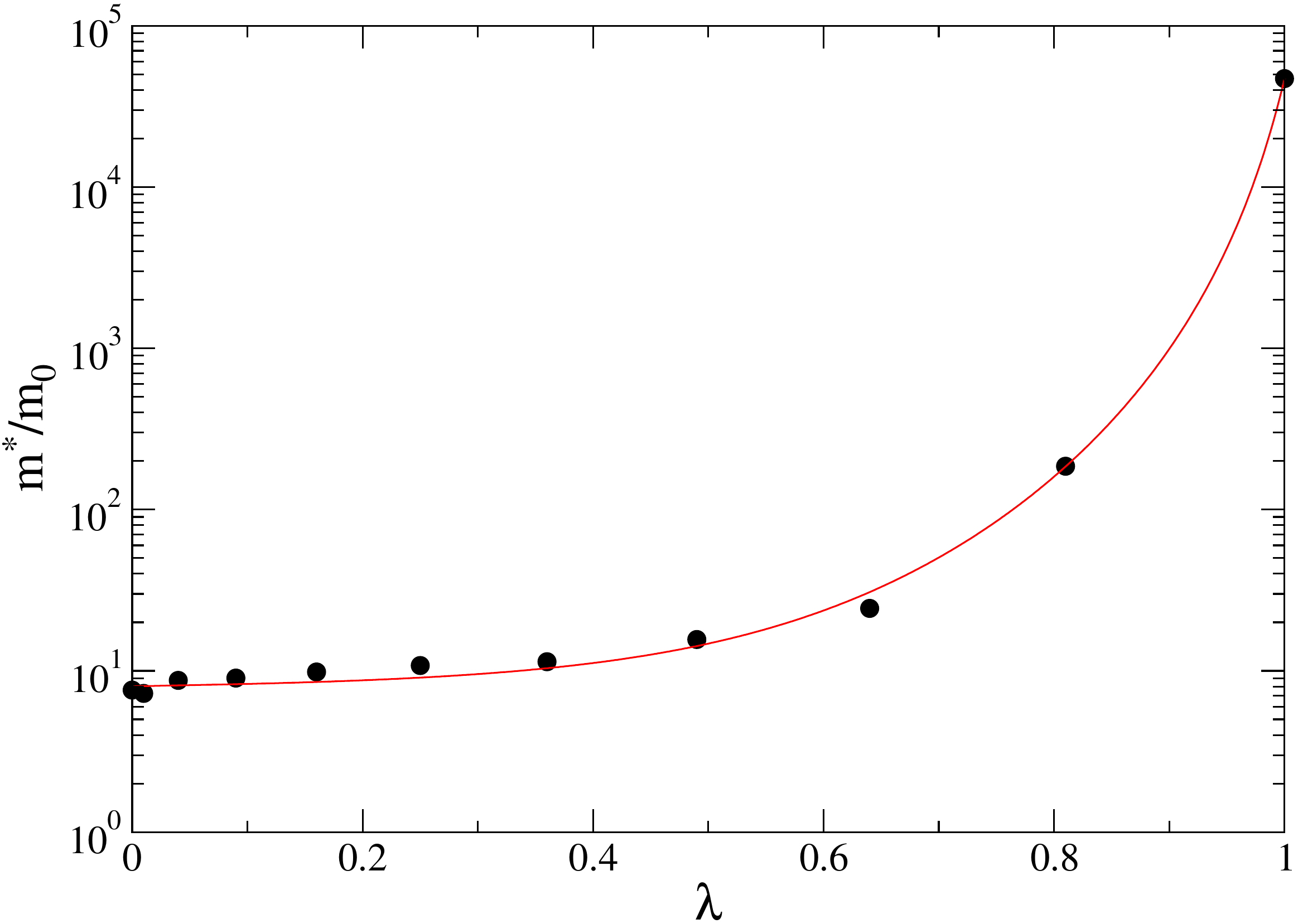}
\end{center}
\caption{(Color online) QP residue $Z$ as  a function of 
electron-phonon coupling $\lambda$ in the KLM with $J=W/2$, $\omega_0=W/2$ while $W=0.2$.
The conduction band filling is  $n_c = 0.8$. The red line is a guide to the eye.  }
\label{NRG_g_scan.fig}  
\end{figure}

We have up to now considered  only a fixed phonon frequency and altered the electron phonon coupling.    
Fig. \ref{QMC_pp_w_omscan.fig}  considers the evolution of $T_{\rm coh}$ at a fixed electron-phonon 
interaction, $\lambda = 0.25$, and as a function of frequency $\omega_0/t$.
As apparent, there is a rapid convergence  to the high-frequency limit.  This
crossover  between the high-frequency and low-frequency behavior  compares remarkably 
well with the energy scale of $T_{\rm K}$ as estimated from the SB
mean-field calculations ({\it cf.} Table~\ref{tab_SB}).

To capture the low energy behavior of the above transition, we concentrate on the KLM of Eq.~(\ref{klm}). This 
model is an effective low energy  model for the PAM in the limit where charge fluctuations on the $f$-sites
are negligible.  Considering the KLM facilitates the study of the  low energy physics in this limit.
We have equally used an NRG solver \cite{Bulla08} to at best resolve the low energies.

In Fig.~\ref{NRG_g_scan.fig} we show  the variation of the effective mass, 
$m^{*}\propto 1/Z$, induced by increasing electron-phonon coupling,  $\lambda$, while keeping the 
phonon frequency fixed at $\omega_0=W/2$.  In analogy  to our results for the PAM (see Fig.~\ref{QMC_pp_w_scan.fig})
one observes  a divergence of $m^{*}$, or equivalently  a vanishing  of the coherence temperature.

\begin{figure}[t!] 
\begin{center}
\includegraphics*[width=0.35\textwidth]{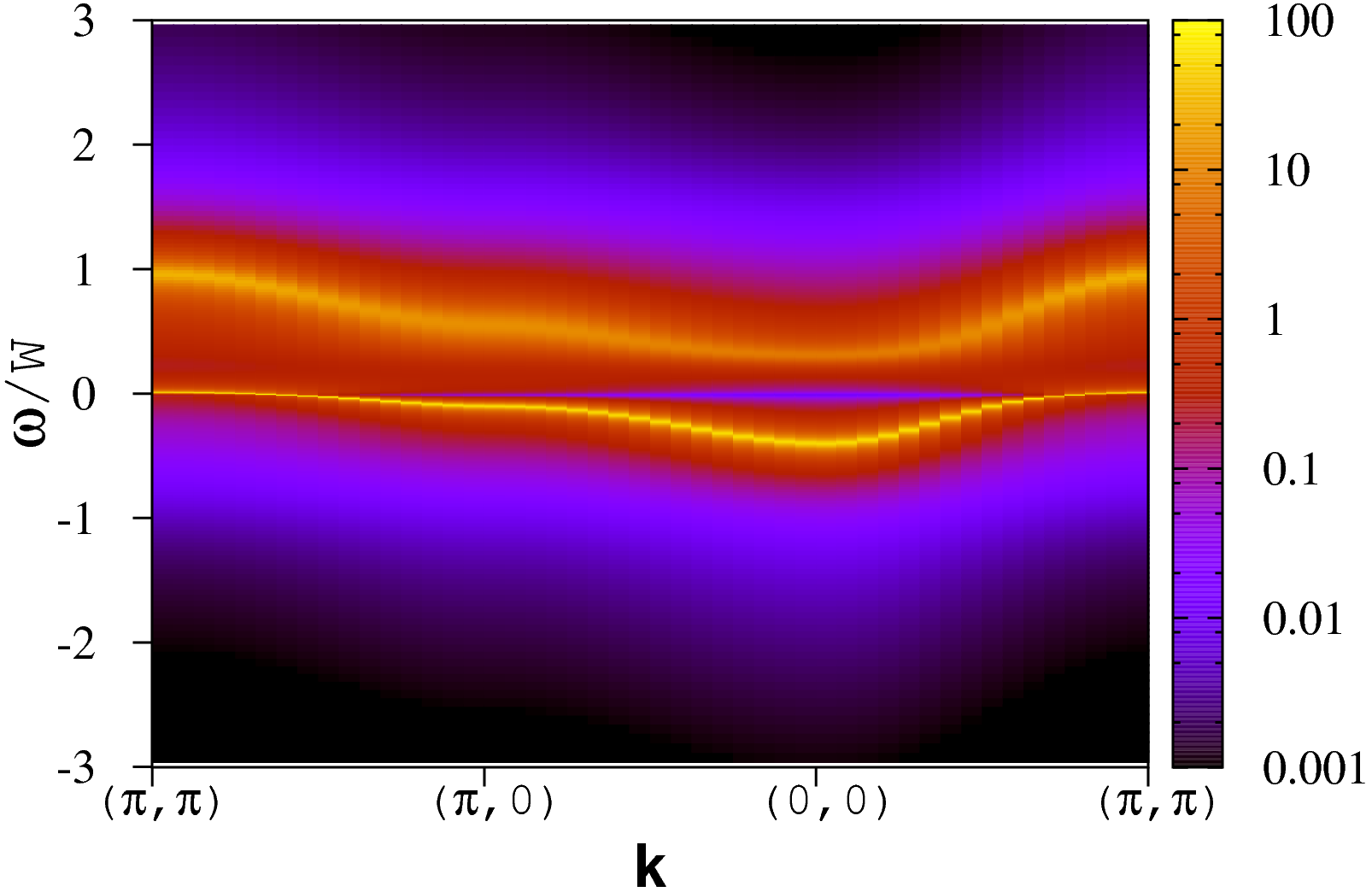} \\
\includegraphics*[width=0.35\textwidth]{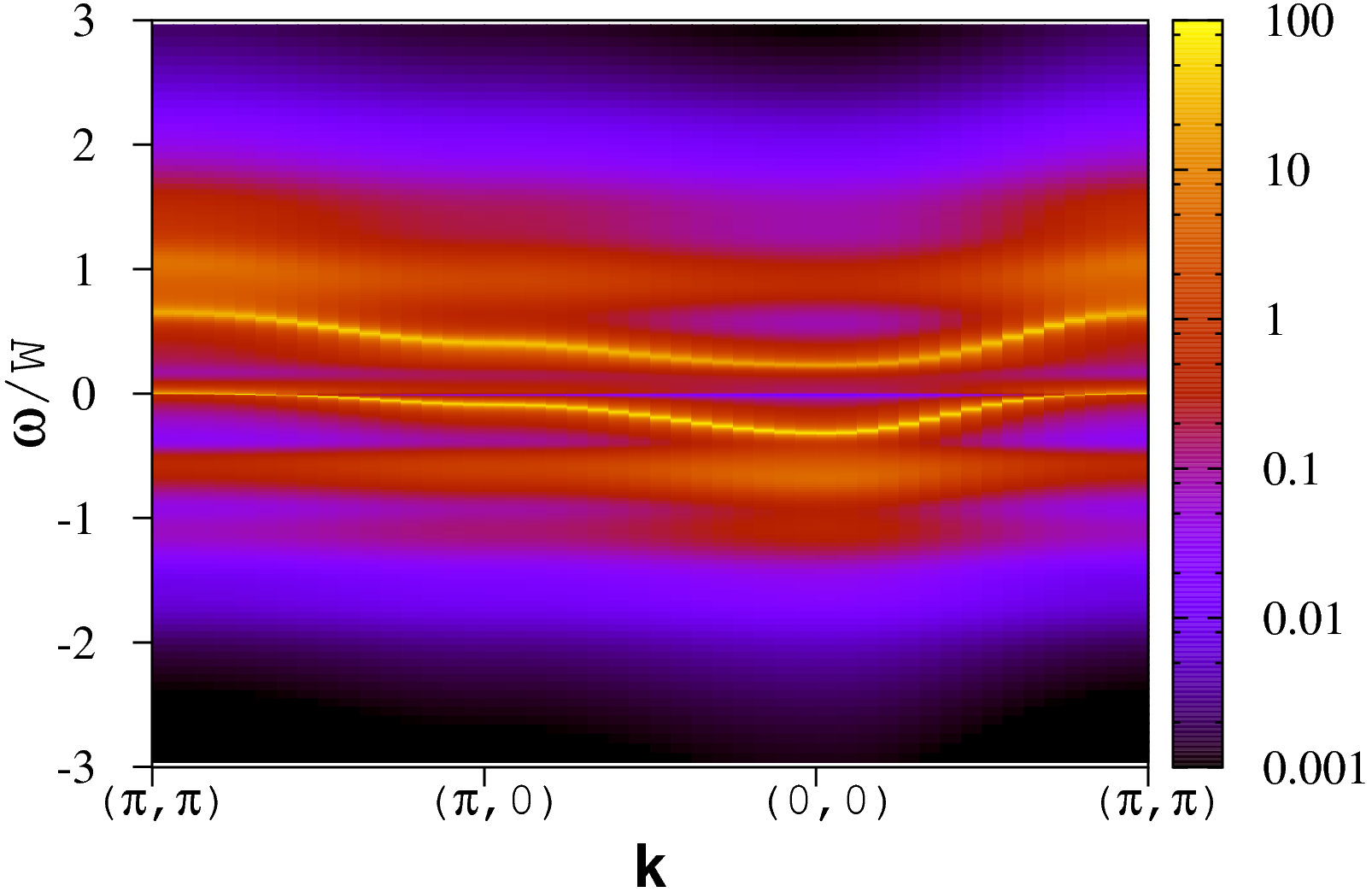} \\
\includegraphics*[width=0.35\textwidth]{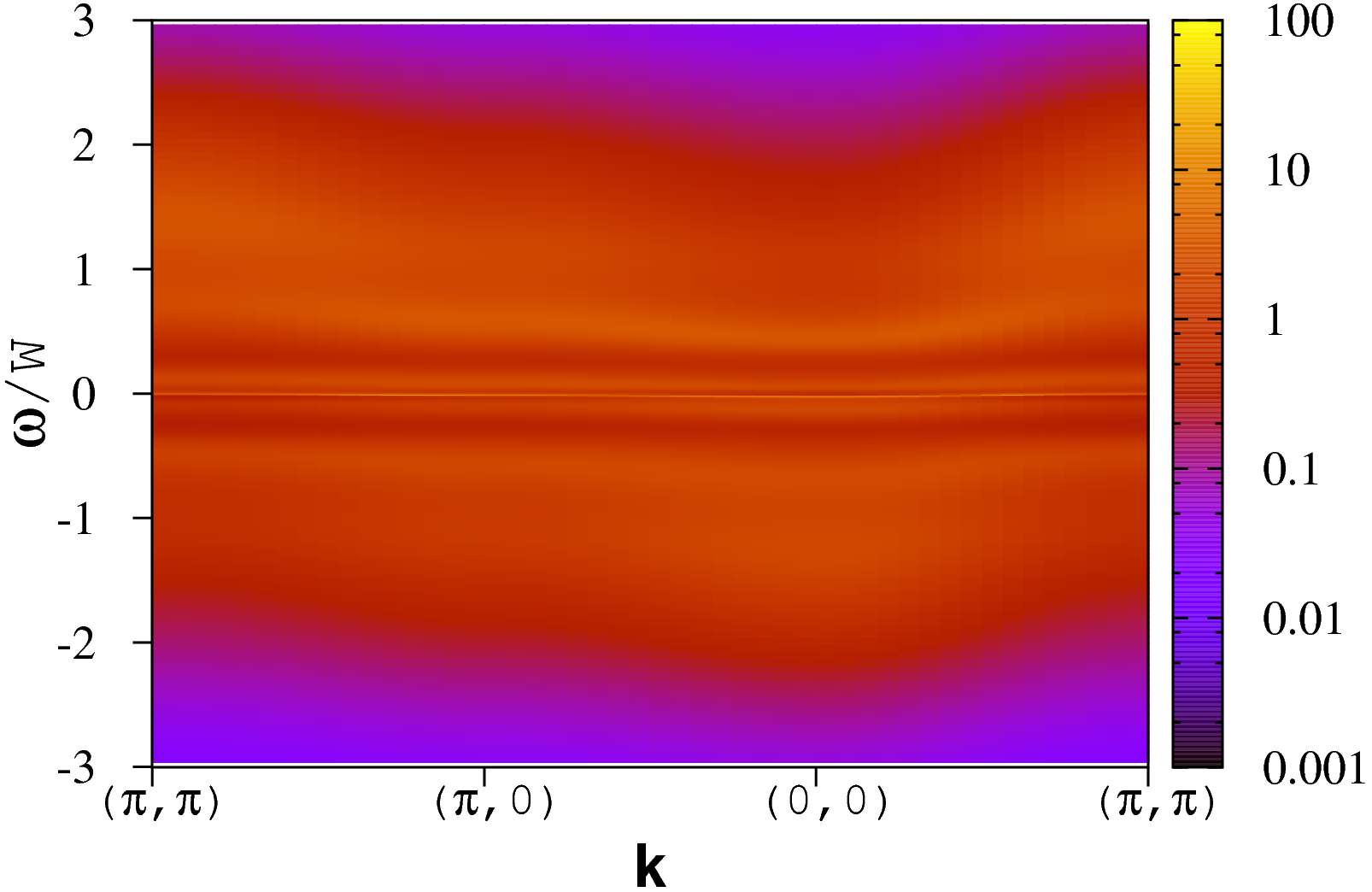} 
\end{center}
\caption{(Color online) Closing hybridization gap on increasing
  electron-phonon coupling $\lambda$ as seen in the 
$c$-electron single-particle spectral function $A_c({\bf k},\omega)$: 
$\lambda=0$ (top), $\lambda=0.25$ (middle), and $\lambda=1$
  (bottom). Parameters as in Fig.~\ref{NRG_g_scan.fig}.
}
\label{NRG_Akom.fig}
\end{figure}

\begin{figure}[t!] 
\begin{center}
\includegraphics*[width=0.45\textwidth]{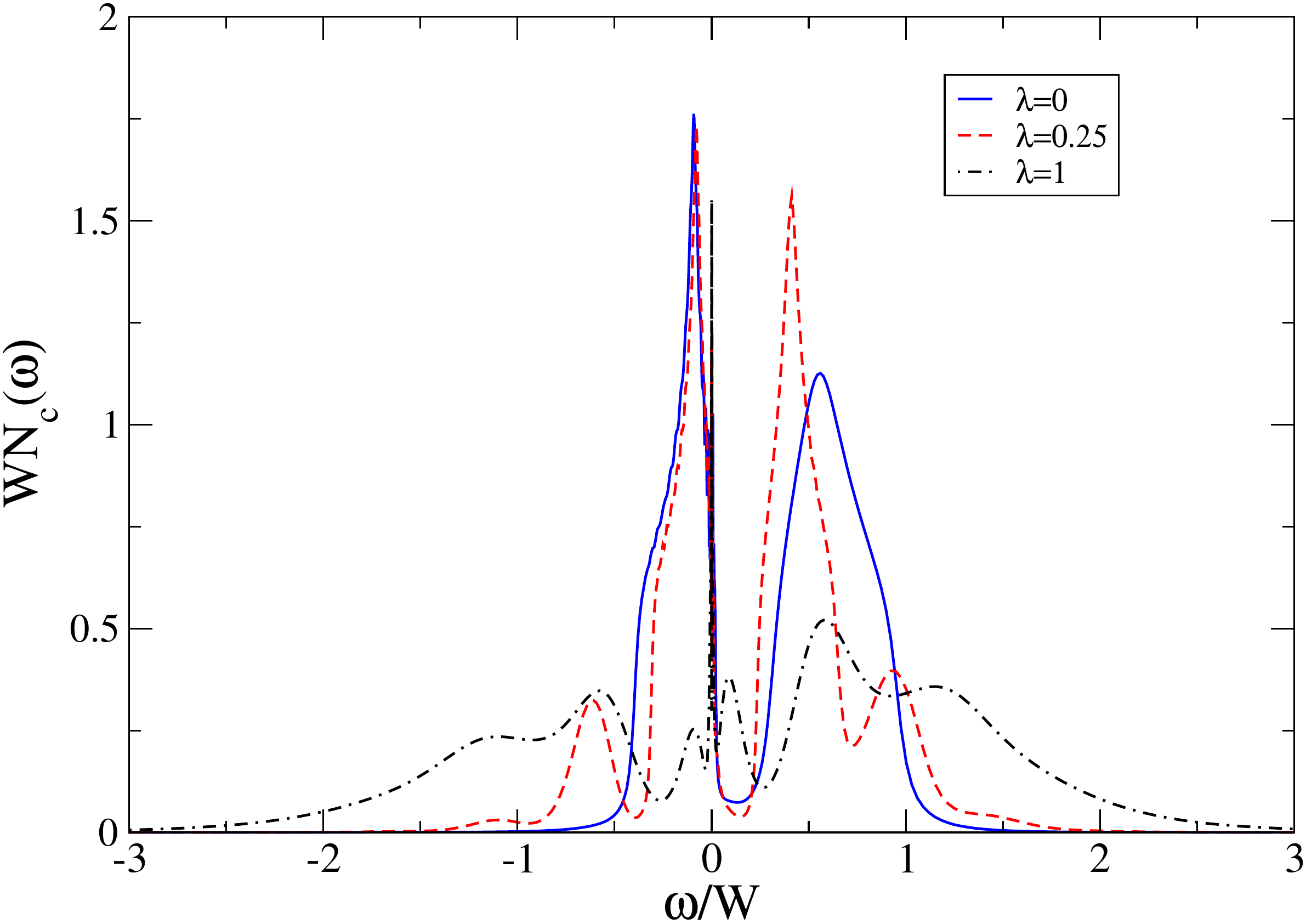}
\end{center}
\caption{(Color online) Closing hybridization gap on increasing
  electron-phonon coupling $\lambda$ as seen in the 
$c$-electron density of states. Parameters as in Fig.~\ref{NRG_g_scan.fig}. 
}
\label{NRG_DOS.fig}  
\end{figure}

Next, it is particularly instructive to follow the single particle spectral function,  Fig.~\ref{NRG_Akom.fig}, 
and corresponding density of states,  Fig.~\ref{NRG_DOS.fig}, as a function of  electron phonon coupling.   
At  weak and intermediate electron-phonon  couplings, $\lambda < 0.8 $,  essentially the same  low energy  single particle 
spectral function is observed   but  with  a renormalized effective mass.  Both at $\lambda = 0$ and $\lambda = 0.25$  the 
single particle density of states  (see Fig.~\ref{NRG_DOS.fig})  shows  a very clear hybridization gap 
which becomes somewhat smaller as a function of  growing electron-phonon coupling thereby reflecting the reduction
of the coherence temperature.

In contrast, the data set at $\lambda = 1$ in Figs.~\ref{NRG_Akom.fig}  and \ref{NRG_DOS.fig}  stands apart.  Here we are in 
the regime where  the dynamics of the $f$-spin is frozen as signalized by $\lim_{T \rightarrow 0} T\chi_s^f   > 0$
[see Eq.~(\ref{Local_suscep.eq})].  In a  static  mean field 
approach,  this state is captured by the vanishing  hybridization matrix element and thereby leads to a total decoupling 
of the $f$- and $c$-electrons. Following this  picture one expects the  conduction-electron spectral function to 
correspond to that of tight-binding electrons coupled to the phonon degrees of freedom thereby producing  coherent 
low energy single particle excitations with enhanced mass. However, this
simple picture does not fit the numerical data which shows no sign of a coherent quasiparticle. Hence, an alternative
interpretation is required. 

The characteristic feature of the state considered here is the frozen dynamics
of the $f$-spin which essentially leads to a separation of time scales; the conduction electron propagates  in  the \emph{static}
magnetic background of the $f$-spins. Since in the DMFT  we do not allow for
spin ordering, it is very tempting  to describe the magnetic background in terms of an $f$-spin 
which points in a random direction on each site.
\begin{figure}[b!] 
\begin{center}
\includegraphics*[width=0.45\textwidth]{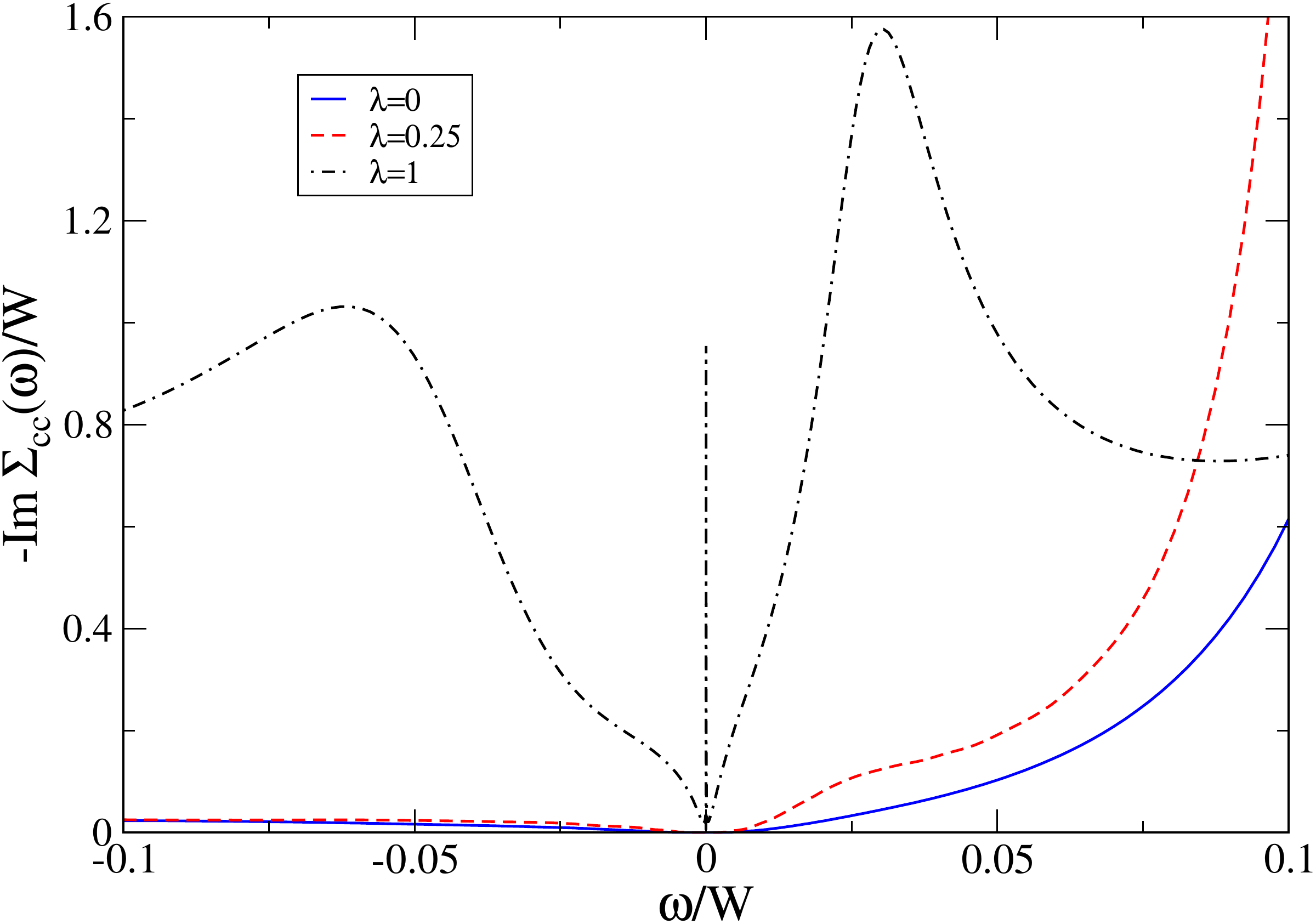}
\end{center}
\caption{(Color online) Low frequency closeup of the imaginary part of the self-energy as obtained for
  increasing electron-phonon coupling from the NRG simulations.  Parameters as in Fig.~\ref{NRG_g_scan.fig}. } 
\label{NRG_ImSig.fig}  
\end{figure} 
This simple model of disorder, which can be dealt with  within 
a Coherent Potential Approximation (CPA) naturally accounts for the incoherent
features of the low energy  spectral function seen at $\lambda = 1$. 
The  incoherence of the single particle spectral function is clearly  seen in the self-energy data of Fig.~\ref{NRG_ImSig.fig}; 
at $\lambda = 1$   and in the low frequency limit, the imaginary part of this quantity remains finite. \cite{note} 
A similar argument was used in Ref.~\onlinecite{Luitz09} to account for  the physics of the single-particle 
spectral function in  the local moment regime of the PAM with $s$-wave superconducting  BCS  conduction electrons.

\section{ \label{sec:3} Conclusions }  

In this paper, we have shown that  the low coherence  temperature characteristic of heavy fermions 
compounds leads to the breakdown of  the Migdal theorem.  In particular, the dominant effect which 
competes  with the formation of the heavy fermion metallic state is conduction electron  binding which inhibits 
Kondo screening of the impurity spins.  Since the Kondo  temperature is a
small scale, already {\it weak} electron-phonon coupling leads to a large reduction of the  coherence temperature and  
ultimately to a Kondo breakdown.  Our studies were conducted in the local moment regime of the periodic
Anderson model and Kondo lattice model in which  phonon degrees of freedom
predominantly couple to the conduction electrons. 
Hence, we can conjecture that heavy fermion materials in the local moment
regime should show no phonon anomalies since this would imply  a
large electron-phonon interaction which as we have shown leads to the breakdown of the heavy fermion metallic state. 
Moreover, since the electron-phonon interaction considerably lowers the
coherence and  Kondo scales, it implicitly enhances the importance of magnetic
instabilities  driven by the Ruderman-Kittel-Kasuya-Yosida (RKKY)
interaction between the local moments.  It also leads to pairing hence potentially 
favoring  superconductivity. The delicate interplay between those phonon driven  instabilities  can only be  understood 
within the framework  of cluster methods and is the goal of future investigations.

\begin{acknowledgments}
The numerical calculations were carried out at the LRZ-M\"unich and
the J\"ulich Supercomputing center. We thank those institutions for their generous allocation of CPU time. 
M.R. acknowledges support from the Alexander von Humboldt Foundation, Polish
Science Foundation (FNP) and from Polish Ministry of Science and Education
under Project No. N202 068 32/1481. F.F.A. thanks the DFG for financial
support. This work was supported in part by the National Science Foundation through OISE-0952300.
\end{acknowledgments}

\appendix* 
\section{Slave-boson technique and the Migdal-Eliashberg approximation}

In the SB method one enlarges the  Hilbert space by introducing a set of four
auxiliary bosonic fields $e_i$, $p_{i\pm\sigma}$, and $d_i$ corresponding to empty,
singly and doubly occupied impurity sites, respectively, such that the $f$-electron operators
are replaced by $f^{}_{i\sigma} \to {\tilde f}_{i\sigma}^{}z_{i\sigma}$, with 
\begin{equation}
z^{}_{i\sigma} = 
\frac{e_{i}^{\dag}p^{}_{i\sigma}+p_{i-\sigma}^{\dag}d_{i}^{}}
{\sqrt{1-d^{\dag}_{i}d_{i}^{}-p_{i\sigma}^{\dag}p_{i\sigma}^{}}
 \sqrt{1-e^{\dag}_{i}e_{i}^{}-p_{i-\sigma}^{\dag}p_{i-\sigma}^{}}}.
\label{eq:z}
\end{equation}
However, such enlargement of the Hilbert space introduces unphysical states 
which should be eliminated so as to recover the original Hilbert
space. Therefore, the SB operators have to fulfill the following constraints 
at each site, 
\begin{equation}
\label{eq:const} 
\begin{aligned}
e_i^{\dag}e_i^{} + d_i^{\dag}d_i^{} 
                  + \sum_{\sigma} p_{i\sigma}^{\dag}p_{i\sigma}^{} &= 1,  \\
d_i^{\dag}d_i^{} + p_{i\sigma}^{\dag}p_{i\sigma}^{} &= {\tilde
  f}_{i\sigma}^{\dag}{\tilde f}_{i\sigma}^{}.
\end{aligned}
\end{equation}

The constraints enforced by time-dependent Lagrange multipliers
$\lambda_i^{(1)}$ and $\lambda_{i\sigma}^{(2)}$ are conveniently handled in a
path integral formulation. Indeed, the SB partition function of the periodic 
Anderson model may be written down as a functional integral over 
coherent states of Fermi and Bose fields,
\begin{equation}
Z_{SB}=\int
D[e,p_{\pm\sigma},d]D[c,{\tilde f}]D[\lambda^{(1)},\lambda^{(2)}_{\pm\sigma}] e^{ -(\mathcal{S}_B + \mathcal{S}_F)},
\label{eq:Z_sb}
\end{equation}
with the bosonic, 
\begin{equation}
\label{eq:bos}
\begin{aligned}
\mathcal{S}_B &= \int\limits_{0}^{\beta}d\tau \sum_i \Bigl \{ 
d_i^{\dag}\bigl(\partial_{\tau}^{} + 
\lambda_i^{(1)}-\sum_{\sigma}\lambda_{i\sigma}^{(2)} + U\bigr)d_{i}^{}
-\lambda_i^{(1)}  \\
&+e_i^{\dag}\big ( \partial_{\tau}^{} + \lambda_i^{(1)} \bigr)e_{i}^{}  
+ \sum_{\sigma} p_{i\sigma}^{\dag}\bigl ( \partial_{\tau}^{} + 
\lambda_i^{(1)}-\lambda_{i\sigma}^{(2)} \bigr)p_{i\sigma}^{}  \Bigr \},
\end{aligned}
\end{equation}
and fermionic,
\begin{equation}
\mathcal{S}_F = \int\limits_{0}^{\beta}d\tau \Bigl\{ 
H_0 + \sum_{i\sigma}{\tilde f}_{i\sigma}^{\dag} 
\bigl(\partial_{\tau}^{}+{\tilde\epsilon_f} \bigr){\tilde f}_{i\sigma}^{}
+ V\sum_{i\sigma}c_{i}^{\dag}z_{i\sigma}^{}{\tilde f}^{}_{i\sigma} \Bigr \},  
\label{eq:fer}
\end{equation}
actions. In Eq.~(\ref{eq:fer}), ${\tilde \epsilon_f} = \epsilon_f +\lambda_{i\sigma}^{(2)} - U/2 - \mu$ 
is a renormalized $f$-level energy. 
Next, we assume the translation and spin SU(2) invariance and apply the
saddle-point approximation in which one replaces all the 
Bose  fields  and Lagrange multipliers by their time-independent averages, i.e., 
$p\equiv\langle p_{i\sigma}^{\dag}(\tau)\rangle =\langle p_{i\sigma}^{}(\tau)\rangle$,
and so on. The site-normalized SB mean-field free energy becomes,
\begin{equation}
\begin{aligned}
F_{SB}  & =  -  \frac{2}{\beta N}\sum_{{\pmb k},n=\pm} 
       \ln\bigl( 1+ e^{-\beta E_{n}({\pmb k})}\bigr) + \mu\langle n \rangle  \\
       & + \lambda^{(1)}(e^2 + 2p_{}^2 + d_{}^2-1)  
        - 2\lambda_{}^{(2)}(p_{}^2+ d_{}^2) +  Ud_{}^2,  
\label{eq:F_sb}
\end{aligned}
\end{equation}
with,
\begin{equation} 
E_{\pm}({\pmb k})= 
\frac{1}{2} \Bigl[ {\tilde \epsilon_f} +  \epsilon({\pmb k}) 
            \pm\sqrt{[{\tilde \epsilon_f} - \epsilon({\pmb k})]^2 +4(zV)^2 } \Bigr],
\label{Epm.eq}
\end{equation}
being the energies of the hybridized bands. 
The equilibrium values of the classical field amplitudes as well 
as of the chemical potential $\mu$ are determined by the minimization $F_{SB}$ 
with respect to these parameters:
\begin{equation}
\label{eq:der_sb}
\begin{aligned}
 \frac{\partial F_{SB}}{\partial\lambda_{}^{(1)}} 
&= e_{}^{2} + 2 p_{}^2 + d_{}^2 -1 = 0, \\
   \frac{\partial F_{SB}}{\partial\lambda_{}^{(2)}} 
&= \langle {\tilde f}^{\dag}_{}{\tilde f}^{}_{}\rangle 
 - \bigl(p_{}^2+ d_{}^2\bigr) = 0, \\
\frac{\partial F_{SB}}{\partial e_{}^{}} 
&= \lambda_{}^{(1)}e_{}^{} + V \frac{\partial
  z_{}}{\partial e_{}^{}} \chi_{cf}  = 0 , \\
\frac{\partial F_{SB}}{\partial d_{}} 
&= \bigl(\lambda_{}^{(1)} - 2\lambda_{}^{(2)} + U\bigr)d_{}^{} +
V\frac{\partial
  z_{}}{\partial d_{}^{}} \chi_{cf}  =0 , \\ 
\frac{\partial F_{SB}}{\partial p_{}} 
&= \bigl(\lambda_{}^{(1)} - \lambda_{}^{(2)}\bigr)p_{}^{}  +
V \frac{\partial
  z_{}}{\partial p_{}^{}} \chi_{cf} =0 , \\ 
\frac{\partial F_{SB}}{\partial\mu} 
&= \langle n \rangle - 2\Bigl( 
 \langle c^{\dag}_{}c^{}_{}\rangle +
\langle {\tilde f}^{\dag}_{}{\tilde f}^{}_{}\rangle\Bigr),
\end{aligned}
\end{equation} 
where we have defined average bond hopping $\chi_{cf}=\langle
c^{\dag}_{}{\tilde f}^{}_{}\rangle + h.c$.
The expectation values $\langle \alpha^{\dag}_{}\beta^{}_{}\rangle$
with $\alpha_{}\equiv\{c_{},{\tilde f_{}}\}$ are obtained from the
local one-particle Green's functions which allows one to combine the SB approach 
with the ME approximation to account for the phonon degrees of freedom. 
The self-consistent evaluation of the diagram of Fig. \ref{SCB.fig}  amounts to solving:   
\begin{equation}
\label{SCB}
        \Sigma_{cc}( i \omega_m )  =  
        \frac{g^{2} \omega_0}{2k} \frac{1}{\beta N} \sum_{ { \pmb k }, i \Omega_m}    
	D^0(i  \Omega_m) G_{cc}( {\pmb k}, i \omega_m - i \Omega_m),
\end{equation}
with 
\begin{equation}
\label{Dyson}
\begin{aligned}
	{ \pmb {\cal G} } ({\pmb k}, i \omega_m)  &\equiv
\left(\begin{array}{cc} 
     G_{cc}({\pmb k}, i \omega_m) & G_{cf}({\pmb k}, i \omega_m)	 \\
     G_{fc}({\pmb k}, i \omega_m) & G_{ff}({\pmb k}, i \omega_m)	
\end{array} \right)  \\
 &= \frac{1} 
         { { \pmb {\cal G} }^{-1}_0({\pmb k}, i \omega_m) - 
         \left(\begin{array}{cc} 
         \Sigma_{cc}(i \omega_m)   & 0 \\ 
                     0             & 0
         \end{array} \right)}.
\end{aligned}
\end{equation}
Here $D^0(i  \Omega_m) = \frac{1}{\omega_0 + i\Omega_m}  + \frac{1}{\omega_0 - i\Omega_m}$ is the bare phonon propagator, 
 $\omega_m$ ($\Omega_m$) are  fermionic (bosonic) Matsubara frequencies, respectively,  and 
$ { \pmb {\cal G} }_0({\pmb k}, i \omega_m) $   is the $2\times 2$ Green function matrix as obtained from the SB  
mean-field Hamiltonian  at the given iteration step. 
Since at a given iteration  we do not have at hand the pole structure 
of $G_{cc}( {\pmb k}, i \omega_m)$ in the complex frequency plane,  we solve the 
above equation  numerically for real frequencies. Namely, we use the  spectral 
representation of the Green's function: 
\begin{equation}
	G_{cc}({\pmb k}, i \omega_m) = \int {\rm d} \omega' \frac{ A_{c}({\pmb k},\omega') }
	                   { i \omega_m - \omega' },
\end{equation}
where $ A_{c}({\pmb k},\omega') = - \frac{1}{\pi} {\rm Im} G_{cc}^{{\rm
    ret}}({\pmb k},  \omega' ) $, 
and perform summation over bosonic Matsubara frequencies so that the
self-energy  reads:
\begin{align}
	\Sigma_{cc}( i \omega_m ) & =  \frac{g^2}{2k} \omega_0 
     \int{\rm d} \omega' N_{c}(\omega') \nonumber \\
 & \times\left\{ \frac{n_b(\omega_0) + 1 - n_f(\omega') } { - \omega' - \omega_0 + i \omega_m} 
 + \frac{n_b(\omega_0) + n_f(\omega') } { - \omega' + \omega_0 + i \omega_m}
 \right\},
\label{sigma_ph}
\end{align}
where $N_{c}(\omega') \equiv \frac{1}{N} \sum_{ {\pmb k} } A_{c}({\pmb
  k},\omega')$ is $c$-electron density of states, while $n_f[\epsilon(\pmb k)] =
\frac{1}{e^{\beta\epsilon(\pmb k)} + 1}$ and  $ n_b(\omega_0) = \frac{1}{e^{\beta
    \omega_0} - 1}$  are the Fermi function and Bose-Einstein distribution, respectively. Hence,
at a given iteration step at which $N_c(\omega)$ is known we  compute with the above equation 
the self-energy on the real frequency axis ($i\omega_m \rightarrow \omega + i\delta $) and thereby 
recompute the single-particle Green's function and corresponding $N_c(\omega)$
until convergence of $\Sigma_{cc}( i \omega_m )$ is reached. This in
turn  allows one to determine the local Green's functions which 
satisfy the usual Dyson equation (\ref{Dyson}).  Explicitly one finds: 
\begin{equation}
\begin{aligned}
G_{cc} ({\pmb k}, i \omega_m) &= 
\frac{1}{i \omega_m - \epsilon({\pmb k})  - \Sigma_{cc}^{eff}(i\omega_m)},  \\ 
G_{ff} ({\pmb k}, i \omega_m) &= 
\frac{1}{i\omega_m - {\tilde\epsilon_f}  - \Sigma_{ff}^{eff}({\pmb k},i\omega_m)}, \\
G_{cf} ({\pmb k}, i \omega_m) &= 
\frac{zV}{ (zV)^2 -\Sigma_{cf}^{eff}({\pmb k},i\omega_m)},
\end{aligned}
\end{equation}
where we have defined the effective self-energies:
\begin{equation}
\begin{aligned}
\Sigma_{cc}^{eff}(i\omega_m) &= \Sigma_{cc}^{}(i\omega_m) +
  \frac{(zV)^2}{i\omega_m - {\tilde\epsilon_f} }, \\
\Sigma_{ff}^{eff}({\pmb k},i\omega_m) &= 
  \frac{(zV)^2}{i\omega_m - \epsilon({\pmb k})  -
    \Sigma_{cc}^{}(i\omega_m)}, \\
\Sigma_{cf}^{eff}({\pmb k},i\omega_m) &= 
[i\omega_m-\epsilon({\pmb k}) - \Sigma_{cc}^{}(i\omega_m)](i\omega_m-{\tilde\epsilon_f} ).
\end{aligned}
\end{equation}
Finally, the  expectation values $\langle \alpha^{\dag}_{}\beta^{}_{}\rangle$ 
entering  the SB saddle-point equations  are obtained by performing the Fourier transformations:
\begin{equation}
\langle \alpha^{\dag}_{}(\tau)\beta^{}_{}\rangle =
-\frac{1}{\beta N} \sum_{{\pmb k},\omega_m} e^{-i\omega_m\tau} G_{\alpha\beta}({\pmb k},i\omega_m),
\end{equation}
and the whole process is iterated until the SB mean-field equations (\ref{eq:der_sb}) are fulfilled.

%\bibliographystyle{./prsty}
%\bibliography{./fassaad,./marcin}

\begin{thebibliography}{10}

\bibitem{lohneysen08}
H. v.~L\"{o}hneysen, A. Rosch, M. Vojta, and P. W\"{o}lfle, Rev. Mod. Phys.
  {\bf 79},  1015  (2007).

\bibitem{Ueda86}
T.~M. Rice and K. Ueda, Phys. Rev. B {\bf 34},  6420  (1986).

\bibitem{Georges00}
S. Burdin, A. Georges, and D.~R. Grempel, Phys. Rev. Lett. {\bf 85},  1048
  (2000).

\bibitem{Pruschke00}
T. Pruschke, R. Bulla, and M. Jarrell, Phys. Rev. B {\bf 61},  12799  (2000).

\bibitem{Assaad04a}
F.~F. Assaad, Phys. Rev. B {\bf 70},  020402(R)  (2004).

\bibitem{Migdal58}
A. Migdal, JETP {\bf 34},  996  (1958).

\bibitem{Ono05}
K. Mitsumoto and Y. Ono, 
Physica C {\bf 426}, 330  (2005); {\it ibid.\/} J. Magn. Magn.  Mater. {\bf 310}, 419  (2007).  

\bibitem{Kotliar86}
G. Kotliar and A.~E. Ruckenstein, Phys. Rev. Lett. {\bf 57},  1362  (1986).

\bibitem{Eliashberg60}
G. Eliashberg, JETP {\bf 11},  696  (1960).

\bibitem{Engelsberg63}
S. Engelsberg and J.~R. Schrieffer, Phys. Rev. {\bf 131},  993  (1963).

\bibitem{Rubtsov05}
A.~N. Rubtsov, V.~V. Savkin, and A.~I. Lichtenstein, Phys. Rev. B {\bf 72},
  035122  (2005).

\bibitem{Assaad07}
F.~F. Assaad and T.~C. Lang, Phys. Rev. B {\bf 76},  035116  (2007).

\bibitem{Assaad08}
F.~F. Assaad, Phys. Rev. B {\bf 78},  155124  (2008).

\bibitem{Bulla08}
R. Bulla, T.~A. Costi, and T. Pruschke, Rev. Mod. Phys. {\bf 80},  395  (2008).

\bibitem{Raczkowski06}
M. Raczkowski, R. Fr\'esard, and A.~M. Ole\'s, Phys. Rev. B {\bf 73}, 174525 (2006); 
Europhys. Lett. {\bf 76},  128 (2006).

\bibitem{Fresard97}
R. Fr\'esard and G. Kotliar, Phys. Rev. B {\bf 56},  12909  (1997).

\bibitem{pam_mag}
A.~M. Reynolds, D.~M. Edwards, and A.~C. Hewson, J. Phys.: Condens. Matter
{\bf 4},  7589  (1992); V. Dorin and P. Schlottmann, Phys. Rev. B {\bf 46},  10800  (1992);
B. M\"oller and P. W\"olfle, {\it ibid.\/} {\bf 48},  10320  (1993); 
R. Doradzi\'nski and J. Spa\l{}ek, {\it ibid.\/} {\bf 56},  R14239  (1997); 
P. D. Sacramento, J. Apar\'icio, and G. S. Nunes,
J. Phys.: Condens. Matter {\bf 22},   065702  (2010). 

\bibitem{Burdin09}
S. Burdin and V. Zlati\'{c}, Phys. Rev. B {\bf 79},  115139  (2009).

\bibitem{Brunner00b}
M. Brunner, F.~F. Assaad, and A. Muramatsu, Phys. Rev. B {\bf 62},  15 480
  (2000).

\bibitem{Mishchenko01b}
A.~S. Mishchenko, N.~V. Prokof'ev, and B.~V. Svistunov, Phys. Rev. B
  {\bf 64},  033101  (2001).

\bibitem{Luscher06}
A. L\"{u}scher, A. L\"{a}uchli, W. Zheng, and O.~P. Sushkov, Phys. Rev. B {\bf
  73},  155118  (2006).

\bibitem{Martinez91}
G. Martinez and P. Horsch, Phys. Rev. B {\bf 44},  317  (1991).

\bibitem{Ramsak98}
A. Ram\v{s}ak and P. Horsch, Phys. Rev. B {\bf
  57},  4308  (1998).

\bibitem{Slezak06}
C. Slezak, A. Macridin, G.~A. Sawatzky, M. Jarrell, and T.~A. Maier, Phys. Rev.
  B {\bf 73},  205122  (2006).

\bibitem{Gunnarsson06}
O. Gunnarsson and O. R\"{o}sch, Phys. Rev. B {\bf 73},  174521  (2006).

\bibitem{Brink00}
J. van~den Brink, P. Horsch, and A.~M. Ole\'s,
  Phys. Rev. Lett. {\bf 85},  5174  (2000).

\bibitem{Beach04a}
K.~S.~D. Beach, arXiv:cond-mat/0403055  (unpublished).

\bibitem{Beach08}
K.~S.~D. Beach and F.~F. Assaad, Phys. Rev. B {\bf 77},  205123  (2008).

\bibitem{note}
Strictly speaking, at $\lambda =1 $ the effective mass  is very
  large but finite (see Fig.~\ref{NRG_g_scan.fig}). One can check that on a
  scale set by the inverse effective mass  the imaginary part of the self energy vanishes. 

\bibitem{Luitz09}
D. Luitz and F.~F. Assaad, \prb {\bf 81}, 024509 (2010).

\end{thebibliography}
%\input{Main.bbl}

\end{document}